\documentclass[nonblindrev]{informs3-neutral}

\SingleSpacedXI
\usepackage[hidelinks]{hyperref}
\usepackage{enumitem} 
\usepackage[table]{xcolor}
\usepackage{multirow}
\usepackage{microtype}
\usepackage{nicefrac}
\usepackage[ruled,vlined,norelsize]{algorithm2e}
\SetKwComment{tcp}{$\triangleright$\ }{}
\DontPrintSemicolon

\SetCommentSty{mycommfont}

\newcolumntype{M}[1]{>{\centering}m{#1}}

\usepackage{natbib}
 \bibpunct[, ]{(}{)}{,}{a}{}{,}%
 \def\bibfont{\small}%
 \def\newblock{\ }%

\newtheorem{property}{Property}

\renewcommand{\Re}{{\mathbb{R}}}

\newcommand{\cM}{{M}}
\newcommand{\cN}{{\mathcal N}}
\newcommand{\cO}{{\mathcal O}}
\newcommand{\cB}{{\mathcal B}}

\newcommand{\cL}{{\mathcal L}}
\newcommand{\tT}{\text{T}}
\newcommand{\tTT}{\text{T}^{\textsc{lb}}}

\TheoremsNumberedThrough     

\EquationsNumberedThrough    

\begin{document}


\RUNAUTHOR{Vidal et al.}

\RUNTITLE{Arc Routing with Time-Dependent Travel Times and Paths}

\TITLE{Arc Routing with Time-Dependent\linebreak Travel Times and Paths}

\ARTICLEAUTHORS{%
\AUTHOR{Thibaut Vidal}
\AFF{Departamento de Informática, Pontif\'{i}cia Universidade Cat\'{o}lica do Rio de Janeiro, \EMAIL{vidalt@inf.puc-rio.br}}
\AUTHOR{Rafael Martinelli}
\AFF{Departamento de Engenharia Industrial, Pontif\'{i}cia Universidade Cat\'{o}lica do Rio de Janeiro, \EMAIL{martinelli@puc-rio.br}}
\AUTHOR{Tuan Anh Pham}
\AFF{ORLab, VNU University of Engineering and Technology, Vietnam \\
Military Logistics Research Center, Military Logistics Academy, Vietnam, \EMAIL{ptanh@anvita.com.vn}}
\AUTHOR{Minh Ho\`ang H\`a}
\AFF{ORLab, Faculty of Computer Science, Phenikaa University, Vietnam, \EMAIL{hoang.haminh@phenikaa-uni.edu.vn}}
} 

\ABSTRACT{%
Vehicle routing algorithms usually reformulate the road network into a complete graph in which each arc represents the shortest path between two locations. Studies on time-dependent routing followed this model and therefore defined the speed functions on the complete graph. We argue that this model is often inadequate, in particular for arc routing problems involving services on edges of a road network. To fill this gap, we formally define the time-dependent capacitated arc routing problem (TDCARP), with travel and service speed functions given directly at the network level. Under these assumptions, the quickest path between locations can change over time, leading to a complex problem that challenges the capabilities of current solution methods. We introduce effective algorithms for preprocessing quickest paths in a closed form, efficient data structures for travel time queries during routing optimization, as well as heuristic and exact solution approaches for the TDCARP. Our heuristic uses the hybrid genetic search principle with tailored solution-decoding algorithms and lower bounds for filtering moves. Our branch-and-price algorithm exploits dedicated pricing routines, heuristic dominance rules and completion bounds to find optimal solutions for problem counting up to 75 services. Based on these algorithms, we measure the benefits of time-dependent routing optimization for different levels of travel-speed data accuracy.
}%



\maketitle

%

\section{Introduction}

\label{sec_intro}

Congestion in city centers causes massive losses (400 billion dollars per year in the USA reported in \citealt{Cookson2018}) and a wide range of adverse effects on inhabitants. Transit times can widely vary during peak hours. With the widespread availability of sensors, mobile devices, and large databases of historical events, driving speeds can now be modeled more precisely to better optimize urban transportation in metropolitan areas. As a consequence, the amount of literature on vehicle routing problems considering time-dependent travel times (TDVRP) has quickly increased \citep{Gendreau2015,Cattaruzza2017}. However, we argue that current TDVRP models present major shortcomings.

\subsection{Time-dependent Routing Models are Oversimplified}
\label{intro:litt}

The wide majority of studies on TDVRP use a complete graph representation of the network in which each vertex corresponds to a service or depot location. In the seminal article of \cite{Ichoua2003}, piecewise-constant vehicle speed functions are associated with the arcs of this complete graph to model time-dependent driving speeds. With these assumptions, a unique quickest path is known to exist between any two locations. However, this is a rough approximation of reality. In practice, time-dependent speed profiles are specific to each street or region. In their survey and practitioners poll, \cite{Rincon2018} highlighted that an inadequate management of time-dependent travel times represents the greatest current barrier to the adoption of vehicle routing software. As the capabilities of current solvers improve, it is increasingly important to account for vehicle speeds at the network level and use time-dependent quickest path calculations \citep{Bast2016} to find travel-time \emph{profiles} between service locations. In such conditions, the quickest path between two locations can change over time, but the FIFO property remains valid \citep{Ghiani2014}, i.e., leaving earlier cannot result in a later arrival.

Very few TDVRP studies have considered time-dependent speed functions at the road network level. \cite{Eglese2006} and \cite{Maden2009} first introduced a road timetabling algorithm, called LANTIME, for TDVRPs with travel time information on streets. The performance of the algorithm was evaluated on road network data from South West England, leading to an estimated 7\% savings in fuel. \cite{Huang2017} later proposed mathematical formulations for a TDVRP minimizing greenhouse gas emissions, in which the speed profiles are specific to each edge of the network. Finally, \cite{Jaballah2019} introduced a mathematical formulation and a simulated annealing solution approach for a TDVRP with time-dependent travel times at the network level.

The aforementioned studies focus on vehicle routing (i.e., node routing) scenarios. Yet, various applications for waste collection, road maintenance, and postal deliveries are more faithfully modeled with services on arcs and edges \citep{Corberan2015}. The resulting capacitated arc routing problems (CARP) require to take additional \emph{mode} decisions, which represent the traversal direction of each service edge \citep{Irnich2008,Vidal2017b}. These mode decisions influence the choice of entry and exit points for each service, and therefore the quickest paths. As discussed in \cite{Gendreau2015}, time-dependent travel times have been largely ignored in the CARP literature, a surprising fact given that most related applications occur in urban contexts. Some research exists on mathematical formulations and solution methods for single-vehicle cases \citep{Black2013,Sun2015,Yu2015}, but these methods remain limited to fairly small data sets counting around $25$ services. In a nutshell, critical methodological challenges still need to be overcome to solve multi-vehicle time-dependent arc routing models at scale.

\subsection{Towards a Time-dependent Capacitated Arc Routing Model}
\label{intro:contrib}

We introduce a formal definition of the time-dependent CARP (TDCARP) and propose state-of-the-art metaheuristics and mathematical programming algorithms to solve it. We consider a deterministic setting, assuming that travel speed estimates are available (i.e., based on traffic history) but will also consider, in our computational experiments, scenarios in which this information is inaccurate. Our travel speed functions are defined according to the IGP model of \cite{Ichoua2003}, but they are specific to the edges of the road network. To our knowledge, this is the first extensive study on the CARP with time-dependent travel times at the network level. 

Our metaheuristic is based on the unified hybrid genetic search (UHGS) with implicit service mode selection \citep{Vidal2012b,Vidal2017b}. It relies on an incomplete solution representation, where service modes (edge traversal directions during services in our case) are optimized on the fly during every route evaluation by dynamic programming (DP). Our DP was adapted to account for time-dependent travel times. Moreover, we develop problem-tailored move evaluation lower bounds to filter non-promising moves during the local search. Our exact method is based on branch-cut-and-price (BCP). 
It exploits valid inequalities, stabilization, ng-route relaxation, tailored pricing algorithms ---fast heuristic pricing, exact pricing with heuristic dominance, and exact pricing using completion bounds generated by backward pricing with maximum speed--- as well as strong branching techniques. This method produces, for the first time, optimal solutions for TDCARP instances with up to 75 services. Our solution methods rely on the preprocessing of \emph{continuous} time-dependent quickest paths for all starting times, using a variant of Bellman-Ford algorithm in which continuous PL functions are maintained in closed form and updated through lower-envelope and composition operations. The resulting quickest path functions are maintained in bucket-based data structures to allow $\cO(1)$ travel-times queries during routing solution. As a result, our algorithms do not need to use time discretization or perform numerous path calculations with fixed start times.

We conduct an extensive computational campaign on benchmark instances derived from the classical CARP benchmarks of \cite{Li1996}, \cite{Beullens2003} and \cite{Brandao2008}. We measure the benefits of time-dependent optimization in scenarios in which only approximate speed information is available. Therefore, our study gives a clear estimate of the benefits of time-dependent optimization and the impact of speed data quality. Summarizing, the main contributions of our work are the following.
\begin{itemize}[nosep]
\item[1)] We introduce a model of time-dependent service- and travel-time functions in the context of arc routing.
\item[2)] We design an effective algorithm to obtain the travel-time functions between origin-destination pairs in a closed form, without approximation or discretization, as well as efficient data structures to query these functions in near-$\cO(1)$ elementary operations.
\item[3)] We propose efficient heuristic and exact solution approaches built upon innovative solution components (e.g., solution decoding algorithms, move filters, heuristic dominance, completion bounds), which have been specifically designed to be effective for the TDCARP.
\item[4)] We conduct extensive computational experiments to evaluate the performance of the proposed solution strategies and to measure the impact of time-dependent optimization with accurate or inaccurate speed information.
\end{itemize}


\section{The Time-Dependent Capacitated Arc Routing Problem}
\label{sec_probstate}

Let $G=(V, E, A)$ be an undirected graph, in which $V$ is a set of vertices, $E$ is an edge set, and $A$ is an arc set. Node $0 \in V$ represents a depot, where $m$ vehicles of capacity $Q$ are based and available at time $0$. $E_R \subseteq E$ and $A_R \subseteq E$ represent edges and arcs requiring service, and $n = |E_R| + |A_R|$ represents the total number of services. Each service $u \in E_R \cup A_R$ is characterized by a nonnegative demand~$q_{u}$. Each edge $u \in E_R$ can be serviced in one of its two possible orientations, called \emph{modes} in \cite{Vidal2017b}. In contrast, the orientation of services to arcs is fixed. To formalize this difference, we associate to each service $u$ a mode set $\cM_u$ representing its possible orientations, in such a way that $\cM_u = \{1,2\}$ if $u \in E_R$ and $\cM_u = \{1\}$ otherwise. Each service should be performed exactly once, but any edge or arc of $E \cup A$ can be deadheaded multiple times while traveling in the network. Finally, the arcs and edges are characterized by time-dependent travel and service speed functions. As a consequence, the FIFO property is respected when traveling and servicing (i.e., starting later does not allow to arrive earlier), and waiting is never profitable (such that we can eliminate this possibility). The TDCARP aims to design up to~$m$ routes in such a way that:
\begin{itemize}[nosep]
	\item Each route starts and ends at vertex $0$ (the depot);
	\item Each service is operated once by a single vehicle;
	\item The demand on each route does not exceed the vehicle capacity $Q$;
	\item The duration of each route does not exceed a maximum value of $D$;
	\item The sum of the route durations is minimized.
\end{itemize}

\paragraph{Speed model.} In this work, we define time-dependent speed functions directly at the network level rather than on the complete graph representation. Therefore, given a planning horizon $[0,H]$, we associate to each arc and oriented edge $(i,j) \in A_R \cup E_R$ a distance $d_{ij}$ along with a piecewise-constant speed function \mbox{$v_{ij} : [0,D] \rightarrow \Re^+$} with $h_{ij}$ pieces representing the travel speed on this link as a function of time, i.e., the distance traveled per time unit. Speeds can change over time according to these functions \emph{while the vehicle is traveling} on a link. In these conditions, the FIFO property is known to hold \citep{Orda1990,Ichoua2003}. We also introduce time-dependent travel and service time functions. For any $(i,j) \in A_R \cup E_R$, we define a positive piecewise-constant travel-and-service speed function $\hat{v}_{ij}: [0,D] \rightarrow \Re^+$ with $\hat{h}_{ij}$ pieces representing the distance \emph{travelled and serviced} per time unit. Asymmetric edge speeds are allowed, such that possibly $v_{ij}(t) \neq v_{ji}(t)$ and $\hat{v}_{ij}(t) \neq \hat{v}_{ji}(t)$. Different arcs and edge orientations can therefore have different speed functions and different breakpoints.

Based on these definitions, the arrival time values $\Phi_{ij}(t_i)$ and $\hat{\Phi}_{ij}(t_i)$ at $j$ when departing from $i$ at $t_i$ are calculated as:
\begin{align} 
\Phi_{ij}(t_i) &= \left\{ x \left| \int_{t_i}^{x} v_{ij}(t) \ dt =  d_{ij} \right. \right\} && \text{ when travelling on $(i,j)$} \label{eq:integ1} \\
\hat{\Phi}_{ij}(t_i) &= \left\{ x \left| \int_{t_i}^{x} \hat{v}_{ij}(t) \ dt =  d_{ij} \right. \right\} && \text{ when servicing $(i,j)$.}
\end{align}
In a similar manner, the departure time values $\Phi_{ij}^{-1}(t_j)$ and $\hat{\Phi}_{ij}^{-1}(t_j)$ from $i$ to arrive at $j$ exactly at time $t_j$ are calculated as:
\begin{align} 
\Phi^{-1}_{ij}(t_j) &= \left\{ x \left| \int_{x}^{t_j} v_{ij}(t) \ dt =  d_{ij} \right. \right\} && \text{ when travelling on $(i,j)$}  \label{eq:integ2}  \\
\hat{\Phi}^{-1}_{ij}(t_j) &= \left\{ x \left| \int_{x}^{t_j} \hat{v}_{ij}(t) \ dt =  d_{ij} \right. \right\} && \text{ when servicing $(i,j)$.}
\end{align}

\section{Methodology}
\label{sec_solmethod}

We separate the description of our solution methods in four sub-sections: 1)~the calculation of a closed-form representation for arrival time functions, 2)~the preprocessing of quickest path profiles, 3)~our exact branch-cut-and-price approach for the TDCARP, and 4)~our hybrid genetic search metaheuristic, designed to produce high-quality heuristic solutions in a controlled amount of time.

\subsection{Continuous Travel Time Functions}
\label{sec:buckets}

All solution algorithms for vehicle routing with time-dependent travel times need efficient methods for travel time queries. When travel time information is provided at the network level, computationally-demanding travel time calculation methods can become a major obstacle to the application of routing optimization algorithms \citep{Vidal2019}. 
Two types of queries are recurrent in the solution process. For this reason, they should be performed as quickly as possible:
\begin{itemize}
	\item \emph{Travel and service time queries}: evaluating $\Phi_{ij}(t)$ or $\hat{\Phi}_{ij}(t)$ on an arc $(i,j)$ at a given time $t$;
    \item \emph{Quickest path queries}: evaluating the earliest arrival time $\Psi_{ij}(t)$ at a node $j$ from a node $i$ at a starting time $t$.
\end{itemize}

Most TDVRP algorithms use an iterative algorithm for travel time queries requiring $\cO(h_{ij})$ time (see Appendix~A). To avoid this overhead, we preprocess continuous representations of $\Phi_{ij}$ and $\hat{\Phi}_{ij}$ as closed-form piecewise-linear (PL) functions and develop efficient data structures which allow queries in $\cO(1)$ time in most situations. We first state two important properties of the arrival time functions, and then proceed with a description of our approach.

\begin{property}
\emph{
Functions $\Phi_{ij}$ are piecewise linear, continuous and monotonic.
}
\end{property}

\noindent
\emph{Proof.}
This follows directly from Equation~(\ref{eq:integ1}) given that functions $v_{ij}$ are piecewise-constant, positive and bounded.\hfill$\square$

\begin{property}
\emph{
Let $t_1,\cdots,t_{h_{ij}-1}$ be the breakpoints of function $v_{ij}$. Function $\Phi_{ij}$ has up to $2(h_{ij}-1)$ breakpoints with values $t_1,\cdots,t_{k}, \Phi_{ij}^{-1}(t_l),\cdots, \Phi_{ij}^{-1}(t_{h_{ij}-1})$ where $k = \argmax \{ x \ | \ t_x \leq \Phi^{-1}_{ij}(D) \}$ and $l = \argmin \{ x \ | \ t_x \geq \Phi_{ij}(0) \}$.
}
\label{teo2}
\end{property}

\noindent
\emph{Proof.}
Define $V_{ij}(x) = \int_{0}^{x} v_{ij}(t) \ dt$.
$V_{ij}$ and its inverse $V^{-1}_{ij}$ are non-negative strictly-increasing PL functions. $V_{ij}$ has breakpoints $t_1,\cdots,t_{k}$ where $k = \argmax \{ x \ | \ t_x \leq \Phi^{-1}_{ij}(D) \}$ and $V^{-1}_{ij}$ has breakpoints $V(t_l),\cdots,V(t_{h_{ij}-1})$ where $l = \argmin \{ x \ | \ t_x \geq \Phi_{ij}(0) \}$. Based on Equation~(\ref{eq:integ1}), we obtain $V_{ij}(\Phi_{ij}(t)) - V_{ij}(t) = d_{ij}$ leading to:
\begin{equation}
\Phi_{ij}(t) = V^{-1}_{ij}(V_{ij}(t) + d_{ij}).
\end{equation}
Function $\Phi_{ij}$ is PL as a composition of two PL functions. Breakpoints can occur whenever $t$ is a breakpoint of $V_{ij}(t)$, or whenever $V_{ij}(t) + d_{ij}$ is a breakpoint of $V^{-1}_{ij}$. In the latter case, there exists $k$ such that $V_{ij}(t) + d_{ij} = V_{ij}(t_k)$, i.e., such that $t = \Phi_{ij}^{-1}(t_k)$.\hfill$\square$\\

Theorem \ref{teo2} holds for function $\Phi_{ij}$ as well as $\hat{\Phi}_{ij}$. Its characterization of the breakpoints leads to a simple procedure to generate a closed-form representation of $\Phi_{ij}$ (or $\hat{\Phi}_{ij}$) in $\cO(h^2_{ij})$ time. This procedure, described in Algorithm \ref{building-functions}, is performed once for each link $(i,j) \in A \cup E$ in a preprocessing phase, before the quickest path algorithm and routing solution method.

\begin{figure}[htbp]
\centering
\begin{minipage}{0.85\linewidth}
\begin{algorithm}[H]
\SingleSpacedXI
\caption{Closed-form construction of $\Phi_{ij}$}
\label{building-functions}
$k \gets \argmax \{ x \ | \ t_x \leq \Phi^{-1}_{ij}(D) \}$\;
$l \gets \argmin \{ x \ | \ t_x \geq \Phi_{ij}(0) \}$\; \vspace*{0.35cm}

$A_\textsc{bp} = \varnothing$ \;
$A_\textsc{bp} \gets  (0,\Phi_{ij}(0))$\;
\lFor{$x \in \{1,\cdots,k\}$}
{$A_\textsc{bp} \gets (t_x,\Phi_{ij}(t_x))$\tcp*[f]{$\cO(h_{ij})$ queries to Algo \ref{alg:calPhi}}} 
\lFor{$x \in \{l,\cdots,h_{ij}-1\}$}
{$A_\textsc{bp} \gets (\Phi_{ij}^{-1}(t_x),t_x)$\tcp*[f]{$\cO(h_{ij})$ queries to Algo \ref{alg:calPhiInverse}}} 
$A_\textsc{bp} \gets  (\Phi^{-1}_{ij}(D),D)$\;  \vspace*{0.35cm}

$A_\textsc{Pieces} = \varnothing$ \;
\textsc{Sort and Remove Duplicates}($A_\textsc{bp}$) \tcp*{$\cO(h_{ij} \log h_{ij})$}

\lFor{$x \in \{1,\cdots,\textsc{Size}(A_\textsc{bp})-1\}$}
{
$A_\textsc{Pieces} \gets (A_\textsc{bp}[x],A_\textsc{bp}[x+1])$
}

\Return $A_\textsc{Pieces}$ \;
\end{algorithm}
\end{minipage}
\end{figure}

The result can be stored as a simple array of function pieces, giving indexed access in $\cO(1)$ time if the index of the piece is known, and $\cO(\log h_{ij})$ otherwise by binary search. To further reduce the query complexity, we divide the planning horizon into $B$ buckets and create an auxiliary array whose i$^{th}$ element points towards the function piece that contains time $(i-1)\frac{D}{B}$. Any travel time query from a time $t$ is resolved by comparing the buckets of indices $\lfloor t/B \rfloor$ and $\lceil t/B \rceil$: if both buckets point towards the same piece, then this piece is returned in $\cO(1)$ time, otherwise a binary search is initiated between the pieces and completed in $\cO(\log h_{ij}$) time.

\subsection{Travel Time Profile Queries}
\label{sec_shortest_path}

Quickest path queries constitute an essential building block of vehicle routing and arc routing algorithms applied to real networks. In most applications, a complete $n \times n$ travel-time matrix is preprocessed to guarantee $\cO(1)$ time, cost, or distance queries through the search. Time-dependent travel times defined at the network level greatly increase the complexity of such preprocessing. Indeed, travel times between origin-destination pairs now depend on an additional continuous parameter, the start time, and the quickest paths may change over time \citep{Ghiani2014}. As a consequence, previous studies did not exploit preprocessing and instead relied on time-consuming quickest-path queries during routing solution. This computational overhead drastically limited the number of local-search moves and construction decisions that can be evaluated in a fixed amount of time, such that only simplistic algorithms, based on much fewer local-search iterations, were practical~\citep{Jaballah2019}.

To circumvent this issue, we use a \emph{continuous} preprocessing approach, during which we compute closed-form representations of the arrival time $\Psi_{ij}(t)$ function at each destination~$j$ \emph{for all} departure times $t$ at origin $i$. This effectively avoids the computational overhead of approaches based on iterative travel time queries as well as the memory overhead and imprecision of approaches based on time discretization. Since the FIFO assumption holds in our context, we use a dynamic programming approach that follows the same scheme as the classical Bellman-Ford algorithm, but in which the discrete labels are replaced by continuous PL functions. This procedure, described in Algorithm \ref{alg:shortestpath}, is applied from the depot and every vertex located at the end of a serviced arc or at one extremity of a serviced edge.

\begin{figure}[htbp]
\centering
\begin{minipage}{0.85\linewidth}
\begin{algorithm}[H]
\SingleSpacedXI
\label{alg:shortestpath}
\SetKwFunction{break}{break}
\SetKwFunction{return}{return}
\textbf{for} $j \in V, \Psi'_{ij} = \Psi_{ij} =
\begin{cases}
\text{id} & \text{ if } i = j \\
\infty & \text{ otherwise }
\end{cases}
$\;

$\mathcal{L} \gets \{i\}$ \;
 \While {$\mathcal{L} \neq \varnothing$} 
{
	\For{$(x,y) \in E \cup A: x \in \mathcal{L}$}
    {
	       $\Psi'_{iy} \gets \textsc{LowerEnvelope}(\Psi'_{iy}, \Phi_{xy} \circ \Psi_{ix})$
	}

$\mathcal{L} \gets \varnothing$\;
	\For{$y \in V$}
{
\If{$ \Psi_{iy} \neq \Psi'_{iy}$}
{
$\mathcal{L} \gets \mathcal{L} \cup y$ \;
$\Psi_{iy} \gets \Psi'_{iy}$\;
}
}
} 
 \caption{Quickest path algorithm from $i$ for all starting times}
 \end{algorithm}
 \end{minipage}
 \end{figure}

In Algorithm~\ref{alg:shortestpath}, ``\text{id}'' represents the identity function. Due to the continuous representation of functions $\Psi_{ij}(t)$, each usual label comparison is replaced by a lower envelope operation \citep{Hershberger1989}, and the  ``$\circ$'' operation corresponds to the composition of two PL functions. These operations are directly performed on the closed-form representation of the functions. The total number of lower envelope and composition operations remains limited to $O|V|(|E \cup A|)$ as in Bellman-Ford algorithm, yet the number of function pieces may grow in the worst case as $n^{\Theta(\log n)}$ \citep{Foschini2014}. As discussed in Section \ref{sec:expSP}, the number of pieces obtained in our computational experiments remains sufficiently small to be efficiently computed and stored in a few seconds.

As a result of this algorithm, we obtain the continuous PL functions $\Psi_{ij}$ representing the \emph{value} of the quickest paths from any vertex $i$ at any time $t$, and store them using the bucket data structure described in Section \ref{sec:buckets}. We do not maintain the time-dependent \emph{paths} themselves, since 1) this information is only useful at the end of routing optimization to produce the detailed solution, 2) this would lead to a significant memory overhead and 3) one can simply use a discrete time-dependent quickest path algorithm for fixed departure times between each pair of successive vertices in the final solution to recover the paths.

\subsection{Branch-Cut-and-Price Algorithm}
\label{metho:BCP}

Our branch-and-price algorithm exploits the same set partitioning model as for most vehicle routing problems. The critical differences with other works occur in the definition of the pricing problem, the labeling algorithms designed to solve it, the branching rules, and the selection of valid inequalities.

Let $\Omega$ be the set of all feasible routes for the TDCARP. Each feasible route $r \in \Omega$ is defined as a sequence of required links (oriented services) $((i_1, j_1), (i_2, j_2), \ldots, (i_k, j_k))$ starting and ending at the depot. Equation~\eqref{eq:route-time} permits to calculate the time of each service as well as the route completion time (return to the depot) as $\Phi_r = \Psi_{j_k0}(T(i_k, j_k))$:
\begin{equation}
\begin{cases}
T(i_\ell, j_\ell) = \hat{\Phi}_{i_\ell j_\ell}(\Psi_{j_{\ell-1}i_{\ell}}(T(i_{\ell - 1}, j_{\ell - 1}))) \\
T(i_1, j_1) = \hat{\Phi}_{i_1 j_1}(\Psi_{0i_{\ell}}(0)).
\end{cases}\label{eq:route-time}
\end{equation}

Let $\lambda_r$ be a binary variable taking value $1$ if and only if route $r \in \Omega$ is used in the solution. Moreover, let $a^r_{ij}$ be a binary parameter which takes value $1$ if and only if route $r$ performs service $(i,j)$. The set partitioning formulation of the TDCARP can be written as follows:
\begin{align}
    \min \hspace*{0.4cm} & \sum\limits_{r \in \Omega} \Phi_r \lambda_r   \label{eq:sp-obj}\\
    \text{s.t.} \hspace*{0.4cm} & \sum\limits_{r \in \Omega} \lambda_r = m  \label{eq:sp-1}\\
    &\sum\limits_{r \in \Omega} a^r_{ij} \lambda_r = 1 & \forall (i, j) \in E_R \cup A_R \label{eq:sp-2}\\
    &\lambda_r \in \{0, 1\} & \forall r \in \Omega. \label{eq:sp-3}
\end{align}
Equation \eqref{eq:sp-obj} minimizes the total duration of the routes. Constraint~\eqref{eq:sp-1} sets a limit on the number of available vehicles, and Constraints~\eqref{eq:sp-2} ensure that each service is performed exactly once.

\noindent
\paragraph{Valid Inequalities.}
Firstly, we use a lifting of the classical odd-edge cutset cuts for the CARP \citep{Belenguer1998}.
In its original form, this family of cuts works on any subset $S$ of nodes with an odd number of incident required links ($|\delta_R(S)|$ odd), by imposing the number of incident deadheaded links ($\delta(S)$) to be at least one. \cite{Bartolini2013a} proved that this family of cuts can be lifted by imposing the same condition on the paths between two required links. Let $\widehat{G} = (V, \widehat{A})$ be a complete graph, called the \emph{deadheaded graph}, where $\widehat{A} = \{(i, j) | i \in V \wedge j \in V, i \neq j\}$, and $\hat{b}^r_{ij}$ is a binary parameter taking value $1$ if and only if route $r$ deadheads the path from $i$ to $j$ after servicing a required link finishing at $i$ and about to start a service on a required link at $j$. Then, the lifted odd-edge cutset cuts can be defined as in Equation~(\ref{eq:odd-edge}):
\begin{equation}
\sum\limits_{(i, j) \in \delta(S)} \hat{b}^r_{ij} \lambda_r \geq 1 \quad \quad \forall S \subseteq V \backslash \{0\}, |\delta_R(S)| \mbox{ odd}. \label{eq:odd-edge}
\end{equation}

Secondly, we use a lifting of the classical capacity cuts \citep{Belenguer1998}. This lifting can be obtained by considering another complete graph $\widetilde{G} = (\widetilde{V}, \widetilde{A})$, called the \emph{required graph}, where $\widetilde{V} = E_R \cup A_R \cup \{0\}$ and $\widetilde{A} = \{(u, v) | u \in \widetilde{V} \wedge v \in \widetilde{V}, u \neq v\}$, i.e., a complete graph where the nodes are the required links plus a node representing the depot. We further use the lower bound on the number of vehicles required to service a set $S \subseteq \widetilde{V} \backslash \{0\}$, represented by $k(S)$, and a binary parameter $\tilde{b}^r_{uv}$ which is $1$ if route $r$ services required link $v$ after required link $u$. A capacity cut defined over this graph, as shown in Equation \eqref{eq:cap-cut}, is similar to a capacitated vehicle routing problem (CVRP) capacity cut. These cuts, proposed by \cite{Bartolini2013a}, are a lifting of the original CARP capacity cuts.
\begin{equation}
\sum\limits_{(u, v) \in \delta(\widetilde{S})} \tilde{b}^r_{uv} \lambda_r \geq 2k(S) \quad \quad \forall S \subset \widetilde{V} \backslash \{0\} \label{eq:cap-cut}
\end{equation}

\noindent
\paragraph{Column Generation.} Formulation (\ref{eq:sp-obj}--\ref{eq:sp-3}) has an exponential number of variables and requires column generation (CG). At each iteration of the CG, the linear relaxation of the set partitioning formulation is solved, and possible negative reduced-cost variables are detected by a pricing algorithm and included in the formulation. The process is iterated until no negative reduced costs variable exists. The reduced costs are computed from the dual information associated with the current solution, using the dual variables $\gamma$, $\beta_{ij}$, $\rho_{ij}$ and~$\pi_{uv}$ associated with Constraints \eqref{eq:sp-1}, \eqref{eq:sp-2}, \eqref{eq:odd-edge} and~\eqref{eq:cap-cut}.

Our pricing algorithm is based on dynamic programming. It computes all candidate paths that start at the depot and that can lead to a negative reduced-cost route. For simplicity, the depot will be represented as a fictitious link $(i_0, j_0)$. Any path $P = ((i_0, j_0), (i_1, j_1), (i_2, j_2), \ldots, (i_h, j_h))$ is represented by a label $\cL(P) = (i(P), j(P), q(P), \Phi(P), \xi(P), \Pi(P))$, containing the last link $(i(P), j(P))$, the total load $q(P)$, the arrival time at the last vertex $\Phi(P)$, the cumulated value of the dual variables, calculated as $\xi(P) = \gamma + \sum_{\ell = 1}^h \left(\beta_{i_\ell j_\ell} + \rho_{j_{\ell - 1} i_\ell} + \pi_{(i_{\ell - 1}, j_{\ell - 1})(i_\ell, j_\ell)}\right)$, and the set $\Pi(P)$ of serviced links. An extension from a path $P$ to a link $(i_\ell, j_\ell)$ is feasible if and only if $q(P) + q_{(i_\ell, j_\ell)} \leq Q$, $\hat{\Phi}_{i_\ell j_\ell}(\Psi_{j(P)i_{\ell}}(\Phi(P))) \leq D$ and $(i_\ell, j_\ell) \notin \Pi(P)$. This extension produces a new label given in Equation \eqref{eq:cg-ext}. At the end of the pricing algorithm, we obtain the routes from the paths such that $(i(P), j(P)) = (i_0, j_0)$ and compute their reduced costs as $\bar{c}(P) = \Phi(P) - \xi(P)$.
\begin{equation}
\begin{aligned}
    \cL(P^\prime) = \Big(i_\ell, j_\ell, q(P) + & q_{(i_\ell, j_\ell)}, \hat{\Phi}_{i_\ell j_\ell}(\Psi_{j(P)i_{\ell}}(\Phi(P))), \\
    &\xi(P) + \beta_{i_\ell j_\ell} + \rho_{j(P) i_\ell} + \pi_{(i(P), j(P))(i_\ell, j_\ell)}, \Pi \cup (i_\ell, j_\ell)\Big)
\end{aligned}
    \label{eq:cg-ext}
\end{equation}

The number of paths considered in the pricing algorithm can grow very quickly.
To mitigate this effect, we use the following rule to discard dominated paths.\vspace*{0.3cm}

\noindent
\textbf{Exact dominance:}
Path $P_1$ exactly dominates path $P_2$ if:
\begin{align}
\text{(i) \quad} &j(P_1) = j(P_2) &\text{(iv) \quad} &\xi(P_1) \geq \xi(P_2) \nonumber \\
\text{(ii) \quad} &q(P_1) \leq q(P_2) &\text{(v) \quad} &\Pi(P_1) \subseteq \Pi(P_2) \nonumber \\
\text{(iii) \quad} &\Phi(P_1) \leq \Phi(P_2). \nonumber
\end{align}

This exact dominance rule uses separate comparisons for the arrival time at the last vertex~$\Phi(P)$ and the cumulated value of the dual variables~$\xi(P)$. These two comparisons appear to be necessary to avoid eliminating promising labels. Consider the following example with two paths: $P_1$ with $\Phi(P_1) = 10$ and $\xi(P_1) = 1$, and $P_2$ with $\Phi(P_2) = 20$ and $\xi(P_2) = 10$. Suppose that $j(P_1) = j(P_2)$, $q(P_1) \leq q(P_2)$ and $\Pi(P_1) \subseteq \Pi(P_2)$. In this example, the reduced costs are $\bar{c}(P_1) = 10 - 1 = 9$ and $\bar{c}(P_2) = 20 - 10 = 10$. If we test $\bar{c}(P_1) \leq \bar{c}(P_2)$ instead of Rules (iii) and (iv), path $P_1$ would appear to dominate $P_2$. However, if an extension exists to a link $(i_\ell, j_\ell)$ for which $\beta_{i_\ell j_\ell} + \rho_{j(P_1) i_\ell} + \pi_{(i(P_1), j(P_1))(i_\ell, j_\ell)} = 1$, $\hat{\Phi}_{i_\ell j_\ell}(\Psi_{j(P_1)i_{\ell}}(\Phi(P_1))) = 3$, and $\hat{\Phi}_{i_\ell j_\ell}(\Psi_{j(P_2)i_{\ell}}(\Phi(P_2))) = 1$ (possible due to the time-dependent travel times), then the reduced costs of paths $P_1$ and $P_2$ after the extension become $10 + 3 - (1 + 1) = 11$ and $20 + 1 - (10 + 1) = 10$, respectively. In this case, a promising label would have been discarded.

Due to this additional comparison, the TDCARP dominance rule eliminates much fewer labels than the typical CARP dominance based on simple reduced cost comparison. To improve this behavior, we propose the following heuristic dominance rule:
\vspace*{0.3cm}

\noindent
\textbf{Heuristic dominance:}
Path $P_1$ heuristically dominates path $P_2$ if:
\begin{align}
\text{(i) \quad} &j(P_1) = j(P_2) &\text{(iv) \quad} &\xi(P_1) + \mu (\Phi(P_2) - \Phi(P_1))  \geq \xi(P_2) \nonumber  \\
\text{(ii) \quad} &q(P_1) \leq q(P_2) &\text{(v) \quad} &\Pi(P_1) \subseteq \Pi(P_2)  \nonumber \\
\text{(iii) \quad} &\Phi(P_1) \leq \Phi(P_2).  \nonumber
\end{align}

\noindent
\paragraph{Discussion.} In rule (iv) of the heuristic dominance, parameter $\mu$ is a positive scalar which weights the contribution of an earlier arrival time at the current link to the final reduced cost at the end of the route. Indeed, it makes sense to assume that an earlier arrival time can lead to a smaller reduced cost at the end of the route. However, as highlighted in Appendix \ref{sec:appB}, arrival time differences can expand or contract over a path by an arbitrary amount, making it difficult to estimate the value of a time gain (or a bound thereof) for an incomplete route. Because of this, the pricing algorithm based on this dominance is only a heuristic. We use it in a first CG phase with a factor $\mu = 0.5$ until no negative cost column is found, and then we switch to the exact pricing algorithm with the exact dominance. During our computational experiments, we observed that only one iteration of the exact pricing was needed in $89\%$ of the cases, i.e., only to \emph{prove} that the CG is completed.

\noindent
\paragraph{Relaxing elementarity.} To further reduce the computational time of the pricing algorithm, we use a variant of the \textit{ng}-route relaxation instead of imposing elementarity \citep{Roberti2011}. For each required link, the approach builds a list of the ``closest'' required links considering an estimated distance measure between two links. Given the required links $(i_1, j_1)$ and $(i_2, j_2)$, the estimated distance between them is $\hat{\Phi}_{i_1 j_1}(0) / 2 + \Psi_{j_1 i_2}(0) + \hat{\Phi}_{i_2 j_2}(0) / 2$. The procedure then considers the minimum between the four possible combinations of service directions. For each required link, we keep the four ``closest'' required links, including itself (i.e., the size of the \textit{ng}-sets is set to four).

\noindent
\paragraph{Backward pricing and completion bounds.} It is not possible to use bidirectional pricing in the TDCARP without keeping (and dominating) speed functions on each label \citep{Lera-Romero2020}. However, we can still obtain completion bounds information from a backward pricing algorithm considering the maximum speed for each link. From its dynamic programming matrix, we build a matrix $\cB(i, j, q, t)$ containing the best reduced cost from the backward paths that end before service $(i, j)$, with load up to $Q - (q - q_{(i, j)})$, using time up to $\lceil D - (t - \hat{\Phi}_{ij}^{+})\rceil$, where $\hat{\Phi}_{ij}^{+}$ is the servicing time of $(i, j)$ on maximum speed. We then use this information to fathom forward label $\cL(P)$ if $\bar{c}(P) + \cB(i(P), j(P), q(P), \lfloor t(P) \rfloor) \geq 0$.

\noindent
\paragraph{Stabilization.} The convergence of the column generation is also improved by applying a dual stabilization procedure. In this work, we use a simple approach that uses a stabilization factor $\alpha \in [0, 1[$, as in \cite{Pessoa2010}. The factor is used on each column generation iteration to perform a convex combination on the dual information from the last iteration and the current one. Our procedure starts with a value of $0.9$, which gives a higher weight to the last iteration dual information. When the optimal solution of the pricing algorithm is not a route with a negative reduced cost, the stabilization factor value is decreased in $0.1$, and the pricing algorithm is called again. This procedure is considered until the factor reaches zero.

\noindent
\paragraph{Fast heuristic pricing.} At the beginning of every column generation iteration, we rely on a fast heuristic pricing to quickly generate promising columns. This procedure is also based on dynamic programming, but it only maintains the path with the best reduced cost for every link and load value \citep{Martinelli2014}. Only when this routine fails to find routes with negative reduced cost, the column generation calls the exact pricing, starting with the heuristic dominance, and terminating with the exact dominance.

\noindent
\paragraph{Branch-and-bound.} The column generation method solves the linear relaxation of Formulation (\ref{eq:sp-obj}--\ref{eq:sp-3}). Even with the valid inequalities \eqref{eq:odd-edge} and \eqref{eq:cap-cut}, integer solutions are not guaranteed and a branch-and-bound approach is required. Thus, whenever a feasible solution is found, we test if this solution is a valid integer solution by checking the arcs on the required graph. If any arc variable has a fractional value, we branch. As discussed in \cite{Pecin2019}, testing the integrality on the required graph is sufficient to guarantee the correctness of the solution.

In addition, we use three branching rules, (i) on the deadheaded degree of a vertex $v \in V$, (ii) on the value of an arc $(i, j) \in \widehat{A}$, and (iii) on the value of an arc $(v, w) \in \widetilde{A}$. The first branching rule follows the same reasoning as the odd-edge cutset cuts \eqref{eq:odd-edge}. If a vertex $v \in V$ has an even (or odd) number of incident required links, then the number of incident deadheaded links must also be even (or odd, respectively). The other two branching rules are straightforward. The choice of branching rules is made using strong branching. A subset of candidates with up to 50 elements is built, and for each candidate, the column generation is solved using only the fast heuristic pricing. Let $z_c^L$ and $z_c^R$ be the objective value of the left and right siblings of candidate $c$, respectively. The procedure selects, as in \cite{Rokpe2012}, the candidate with highest ranking based on the formula of Equation~(\ref{eq:sb-rank}) with $\alpha = \nicefrac{3}{4}$.
\begin{equation}
\textsc{Rank}(c) = \alpha \min(z_c^L, z_c^R) + (1 - \alpha) \max(z_c^L, z_c^R)
\label{eq:sb-rank}
\end{equation}
A new branch with two nodes is then created, and the entire column generation is solved, including the valid inequalities. The next node to explore is obtained following the best-bound strategy.

\subsection{Hybrid Genetic Search}
\label{metho:HGS}

As observed in our experiments, our branch-cut-and-price algorithm optimally solves most problem instances with up to 75 services in a few minutes. Still, it fails to solve larger instances due to limited memory and time. Optimal solutions of this scale are invaluable for experimental comparisons and validations, but remain insufficient for most applications. 

To produce solutions for larger instances as well as initial upper bounds, we introduce an efficient population-based metaheuristic. Our algorithm follows the same principles as the Unified Hybrid Genetic Search (UHGS), which effectively combines crossover-based solution generation, local search, and population-diversity management to produce high-quality solutions for numerous vehicle-routing and arc-routing problem variants \citep{Vidal2012b}. The recent application of the UHGS principle to the family of arc-routing problems \citep{Vidal2017b} exploits two other key strategies: 1) an indirect solution representation in the local search, in which each route is represented as a sequence of services, with an exact decoding algorithm to find optimal mode choices, and 2) efficient lower-bounds to filter non-promising moves in amortized $\cO(1)$ time. Our adaptation of this method to the TDCARP requires replacing the solution decoder and lower bounds while retaining the same solution representation and search operators. For the sake of brevity, we concentrate our description on the new solution decoder (Section \ref{subsec:Decoder}) and lower bounds (Section~\ref{subsec:LB}) and refer to \cite{Vidal2017b} for a description of the general search strategy which remains identical. Moreover, an open-source implementation of the complete TDCARP algorithm is provided at \url{https://github.com/vidalt/HGS-TDCARP}.

\subsubsection{Indirect Solution Representation and Search Space.}
\label{subsec:Decoder}

\begin{figure}[htbp]
    \centering
    \includegraphics[scale=0.84]{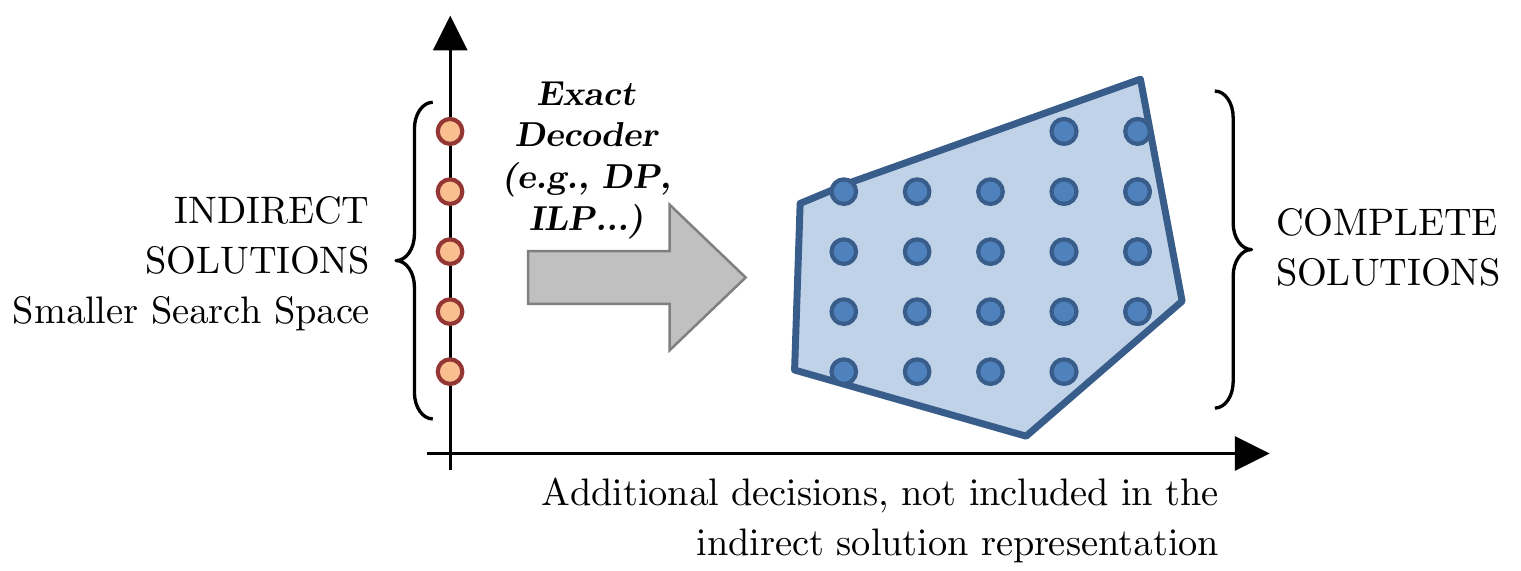}
    \caption{Indirect solution representation and decoding}
    \label{problem-decomposition}
\end{figure}

Indirect solution representations and decoders have been instrumental in designing state-of-the-art solution approaches for optimization problems mixing different decision-variables classes \citep{Vidal2014b,Vidal2014,Vidal2017b,Herszterg2019,Mecler2019,Toffolo2019}.
As illustrated in Figure \ref{problem-decomposition}, the goal of an indirect solution representation is the same as a projection-based problem decomposition. By representing the solution with a smaller subset of decision variables and systematically using an optimal (e.g., dynamic programming-based) decoder to complete the solution, the heuristic searches a smaller space while the rest of the decisions are taken exactly. This representation simplifies the solution approach by making it more structured and greatly reduces the approximations inherent to heuristic search.

The TDCARP combines four classes of decisions: \textsc{Assignment} of services to vehicles, \textsc{Sequencing} of the services within each route, \textsc{Mode} choices for the services (i.e., choices of directions for the services on edges), and \textsc{Path} choices between the services. Each of these decision sets contains a number of options that grows exponentially with the number of services. Moreover, in contrast with the TDVRP where services correspond to nodes in the network, the quickest paths between the extremities of successive edge services in the TDCARP are conditioned by the choices of service orientations, i.e., the \textsc{Mode} decisions.

\begin{figure}[htbp]
\hspace*{-0.27cm}
\renewcommand{\arraystretch}{1.4}
\scalebox{0.918}
{
\begin{tabular}{|M{2.2cm}|m{7.4cm}|m{7.4cm}|}
\hline
&\multicolumn{1}{c|}{\textsc{\underline{R-Complete}}} & \multicolumn{1}{c|}{\textsc{\underline{R-Indirect}}} \\
&\multicolumn{1}{c|}{\textsc{Assignment}, \textsc{Sequences} and \textsc{Modes}} & \multicolumn{1}{c|}{only \textsc{Assignment} and \textsc{Sequences}} \\
\hline
Number of 

solutions in 

search space & \centering $N_\textsc{CVRP} \times 2^{|E_R|}$

$N_\textsc{CVRP}$ representing the number of solutions of a CVRP instance with $n$ customers & 
\multicolumn{1}{c|}{$N_\textsc{CVRP}$} \\
\hline
Decoder
& 
Iterative propagation of time-dependent travel times (example with $k=3$ services):

\includegraphics[scale=0.75]{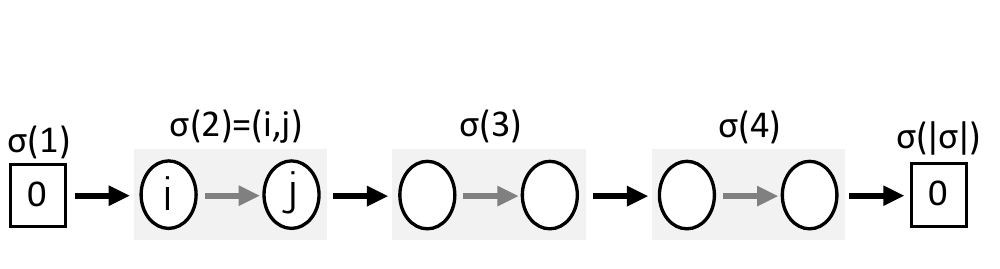}
&
Time-dependent quickest path for a discrete departure date in an acyclic auxiliary graph (example with $k=3$ services to edges):

\includegraphics[scale=0.75]{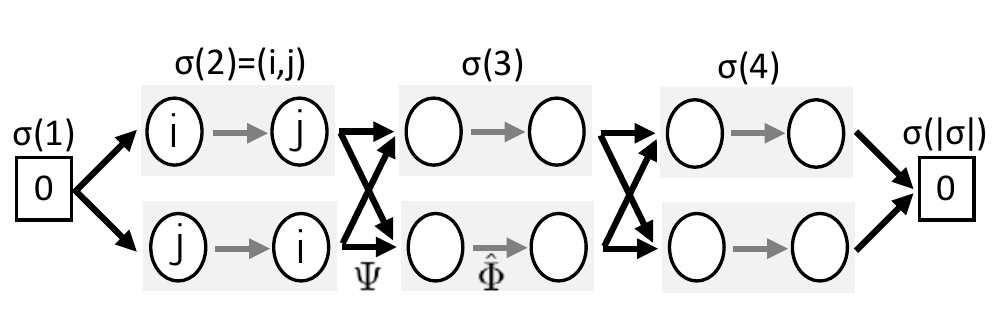}
\\
\hline
\end{tabular}
}
\caption{Solution representations, search spaces and decoders}
\label{comparison-representations}
\end{figure}

To efficiently optimize the \textsc{Mode} decisions, we use the same solution representation as \cite{Vidal2017b}, in which a solution is represented as sequences of services (routes) without their mode information, and systematically use a dynamic programming decoding algorithm to complete the solutions. Figure \ref{comparison-representations} compares this representation choice (\textsc{R-Indirect}) with a classical complete solution representation (\textsc{R-Complete}).

Due to the time-dependent travel times, solution evaluations require to propagate arrival and service completion times over the auxiliary graph illustrated in Figure \ref{comparison-representations} (rightmost side). In \textsc{R-Indirect}, this propagation is done using Bellman's algorithm in topological order, using the information of the quickest paths~$\Psi$ from the preprocessing step to evaluate the travel times between edge extremities (black edges on Figure \ref{comparison-representations}), as well as the service time functions~$\hat{\Phi}$ (gray edges). For a route $\boldsymbol\sigma = (\sigma(1),\dots,\sigma(|\sigma|))$ represented according to \textsc{R-Indirect} as a sequence of services starting and ending at the depot (such that $\sigma(1)=0$ and $\sigma(|\sigma|)=0$), the completion time $T^\textsc{Exact}_{\sigma(i)}[x]$ of each service $\sigma(i)$ for each mode $l \in \cM_{\sigma(i)}$ can be calculated as:
\begin{align}
\tT_{\sigma(i)}^\textsc{Exact}[l] = 
\begin{cases}
0 & \text{if } i=1 \\
\min\limits_{k \in \cM_{\sigma(i-1)}} \{ \hat{\Phi}^l_{\sigma(i)}(\Psi^{kl}_{\sigma(i-1)\sigma(i)}(\tT^\textsc{Exact}_{\sigma(i-1)}[k])) \} & \text{otherwise},
\end{cases}
\label{propagationn}
\end{align}
and the route duration is given by $\tT^\textsc{Exact}_{\sigma(|\sigma|)}[1]$.

In Equation (\ref{propagationn}), $\Psi^{kl}_{ij}(t)$ represents the arrival time when leaving the end of service $i$ in mode~$k$ at time~$t$ towards the origin of service~$j$ in mode~$l$, and $\hat{\Phi}^l_i(t)$ represents the completion time of service $i$ in mode $l$ starting at time $t$. Let $C_{\Psi}$ and $C_{\hat{\Phi}}$ be the complexity of querying the arrival time functions~$\Psi$ and the service time functions~$\hat{\Phi}$, respectively. With this notation, route evaluations in \textsc{R-Indirect} take $4k C_{\Psi} + 4k C_{\hat{\Phi}}$ time, in comparison to $(k+1) C_{\Psi} + k C_{\hat{\Phi}}$ time for \textsc{R-Complete}. This is a fourfold increase of evaluation effort, but it dramatically reduces the size of the search space (by a factor $2^{|E_R|}$) and renders \textsc{Mode} and \textsc{Path} decisions optimal and transparent within the search. As will be discussed in the next section, the computational effort of route evaluations can be further mitigated with move filters based on route-cost lower bounds, and their evaluation by preprocessing and concatenation.

\subsubsection{Efficient Local Search.}
\label{subsec:LB}

Local searches maintain an incumbent solution $s$ and explore a neighborhood $\cN(s)$ obtained by testing some changes on $s$, called moves. The local search of UHGS uses a first improvement policy in which $\cN(s)$ is explored in random order, and any improving move is directly applied. It uses classical intra-route and inter-route moves: \textsc{2-opt}, \textsc{2-opt*}, as well as \textsc{Relocate} and \textsc{Swap} of 0, 1 or 2 consecutive services with possible reversals, subject to proximity restrictions \citep{Laporte2014a,Vidal2012a,Vidal2012b}. We now recall two properties which are fundamental to efficiently filter moves in the TDCARP, and then proceed with the definition of the bounds.

\begin{property}
\emph{
All the considered moves consist of disconnecting up to two routes of the incumbent solution $s$ into a constant number of sequences of consecutive visits and concatenating these sequences in a different order \citep{Vidal2015b}. Preprocessing auxiliary information on the sequences of consecutive visits from $s$ can help to reduce the evaluation complexity of routes obtained from their concatenations.
}
\end{property}

\begin{property}
\emph{
Let $\Pi$ be a move which transforms a pair of routes ($\sigma_1,\sigma_2)$ into $(\sigma'_1,\sigma'_2)$. Let $C_\textsc{lb}(\sigma)$ be a lower bound on the cost of a route $\sigma$. If $C_\textsc{lb}(\sigma'_1) +  C_\textsc{lb}(\sigma'_2) -  C(\sigma_1) -  C(\sigma_2) \geq 0$, then $\Pi$ is non-improving and can be filtered out.
}
\label{propfiltering}
\end{property}

\paragraph{Lower bounds on move evaluations for the TDCARP.}
To efficiently evaluate move lower bounds, our method uses preprocessing and concatenation principles and therefore maintains auxiliary information on the sequences of consecutive services from $s$.
More precisely, any sequence of consecutive visits $\sigma$ is characterized by a lower bound $\tTT(\sigma)[k,l]$ on the travel and service time over $\sigma$, starting the first service in mode~$k$ and finishing the last service in mode~$l$, for each mode combination. This information is calculated by induction. It is
preprocessed at the start of the local search and updated after each change in the solution $s$ using a simple lexicographic calculation method over the services. For a sequence~$\sigma$ containing a single service~$i$, we have:
\begin{equation}
\tTT(\sigma)[k,l] =
\begin{cases}
 \min\limits_{t \in [0,D]} \{ \hat{\Phi}^k_{i}(t)-t \} & \text{if } k = l \\
 \infty & \text{otherwise}.
\end{cases} \label{eqinit}
\end{equation}
In this equation, $\min_{t \in [0,D]} \{ \hat{\Phi}^k_{i}(t) - t \}$ is a scalar which represents the minimum time needed to operate service~$i$ in mode~$k$. This value can be easily collected from the closed-form service-time function and subsequently accessed in~$\cO(1)$. Now, these bounds can be evaluated for larger sequences of services $\sigma_1 \oplus \sigma_2$ arising from the concatenation of any two sequences~$\sigma_1$ and~$\sigma_2$:
\begin{equation}
\tTT(\sigma_1 \oplus \sigma_2)[k,l] = 
\min_{x,y} 
\left\{ \tTT(\sigma_1)[k,x] + \min_{t \in [0,D]} \{\Psi^{xy}_{\sigma_1(|\sigma_1|)\sigma_2(1)}(t)-t\} + \tTT(\sigma_2)[y,l]\right\}. \label{eqconcat}
\end{equation}
Equation (\ref{eqconcat}) provides a lower bound on the shortest time needed to perform $\sigma_1$ followed by $\sigma_2$, starting and finishing in modes~$k$ and~$l$ respectively. This equation is used to preprocess the data (by iteratively appending one node) and to evaluate the moves as a concatenation of existing sequences. It is important to remark that $\min_{t \in [0,D]} \{ \Psi^{xy}_{ij}(t) - t \}$ used in this equation represents the value of the quickest path \emph{at the best starting time $t$} between the end extremity of service~$i$ in mode~$x$ and the starting extremity of service~$j$ in mode~$y$.
This value is obtained as a by-product of the calculation of the continuous quickest-path functions $\Psi$ (prior to the routing optimization phase, as discussed in Section~\ref{sec_shortest_path}). It gives a much tighter travel-time bound than a discrete shortest path algorithm on the network that fixes arc and edge lengths to their minimum travel time values over the planning horizon.

Finally, we can strengthen our bound by remarking that, in a route $\sigma_1 \oplus \dots \oplus \sigma_S$ obtained by the concatenation of $S$ sequences, the exact arrival time values (without any approximation) over the first sequence are known from the current incumbent solution~$s$. The resulting lower bound is:
\begin{align}
\tT^\textsc{lb+}(\sigma_1 \oplus \dots \oplus \sigma_S) &= 
\min_{x,y}
\left\{\Psi^{xy}_{\sigma_1(|\sigma_1|)\sigma_2(1)}(\tT^\textsc{Exact}_{\sigma_1(|\sigma_1|)}[x]) + \tTT(\sigma_2  \oplus \dots \oplus \sigma_S)[y,1] \right\}
\label{finalLBmove}
\end{align}
For a route composed of $S$ sequences, calculating this lower bound requires four evaluations of the travel time functions $\Psi$ as well as a constant number of additions proportional to $S$. The lower bound evaluation is thus one order of magnitude faster than the exact move evaluation. Our algorithm first evaluates this lower bound for each route involved in a move and possibly rejects this move based on Property \ref{propfiltering}, otherwise it subsequently uses Equation (\ref{propagationn}) for an exact time evaluation. 
In our experiments, this strategy permits to filter 91\% of the moves, to such an extent that exact move evaluations do not represent anymore a search bottleneck.

\section{Computational Experiments}
\label{sec_experiment}

We conduct extensive computational experiments with two main objectives.
\begin{itemize}[nosep]
\item[1)] We evaluate the performance of the proposed algorithms as well as the effectiveness of some of their principal components: the continuous quickest-path procedure, the pricing using heuristic dominance and completion bounds in the BCP, as well as the efficient structures for travel time queries and lower bounds used to filter local-search moves in HGS.
\item[2)] We measure the impact of time-dependent routing and path optimization, considering different levels of speed inaccuracies.
\end{itemize}

\subsection{Experimental Setup and Benchmark Instances}
\label{sec:expData}

All methods from this paper were implemented in C++, using CPLEX 12.8 for the solution of the master problems and linear programs in our BCP algorithm. The tests were conducted on a single thread of an Intel Core i7-8700K 3.70GHz CPU.

For our analyses, we generated three classes of TDCARP benchmark instances with different levels of time dependency. The network information, customer demands, and vehicle capacities were extracted from the first ten BMCV data sets (C01--C10, from \citealt{Beullens2003}) built from Flanders road network, and from ten EGL data sets \citep{Li1996,Brandao2008} associated with a winter gritting case study in Lancashire. For each of these 20 data sets, we generated three instances with different degrees of time dependency: low~(L), medium~(M), and high~(H). In each case, the piecewise-constant speed profile $v_{ij}(t)$ of each link $(i,j)$ was generated by randomly selecting six distinct breakpoints in $\left\{0.05 D,0.1 D, 0.15D,\dots,0.95 D\right\}$, and then randomly selecting a speed value for each function segment $k \in \{1,\dots, 7\}$ from a uniform distribution $U(a_k, b_k)$ parametrized as indicated in Table~\ref{parameters-speeds}. Finally, we assumed that travel-and-service speed represents 70\% of deadheading speed, i.e., $\hat{v}_{ij}(t) = 0.7 \times v_{ij}(t)$ for all $(i,j) \in E \cup A$, and used the same fleet-size limit as the original CARP data sets.

\begin{table}[htbp]
\renewcommand{\arraystretch}{1.25}
\setlength{\tabcolsep}{8pt}
\begin{center}
\scalebox{0.9}
{
\begin{tabular}{|c|c|c|c|}
\hline
k & Type L & Type M  & Type H \\
\hline
1 & [0.6,0.9] & [0.5,0.8] & [0.4,0.7] \\
2 & [0.8,1.0] & [0.7,1.0] & [0.6,1.0] \\
3 & [1.0,1.3] & [1.0,1.4] & [1.0,1.6] \\
4 & [0.9,1.1] & [0.8,1.2] & [0.7,1.3] \\
5 & [1.0,1.3] & [1.0,1.4] & [1.0,1.6] \\
6 & [0.8,1.0] & [0.7,1.0] & [0.6,1.0] \\
7 & [0.6,0.9] & [0.5,0.8] & [0.4,0.7] \\
\hline
\end{tabular}}
\end{center}
\caption{Parameters $[a_k,b_k]$ for the generation of the speed scenarios}\label{parameters-speeds}
\end{table}

\subsection{Performance of the Time-dependent Quickest Path Algorithm}
\label{sec:expSP}

\begin{figure}[htbp]
    \hspace*{-0.7cm}
    \begin{minipage}{8.5cm}
    \includegraphics[width = \textwidth]{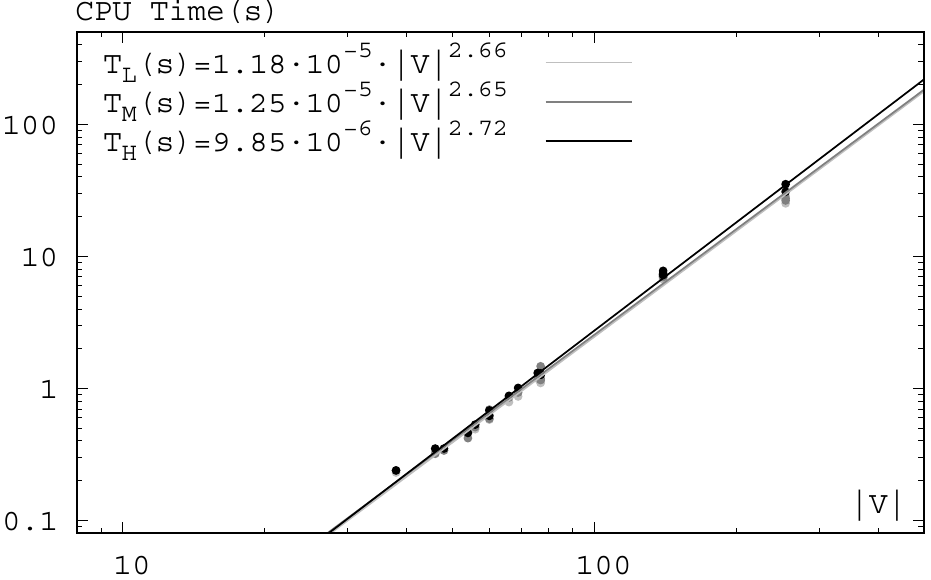}
    \end{minipage}
    \begin{minipage}{8.5cm}
    \includegraphics[width = \textwidth]{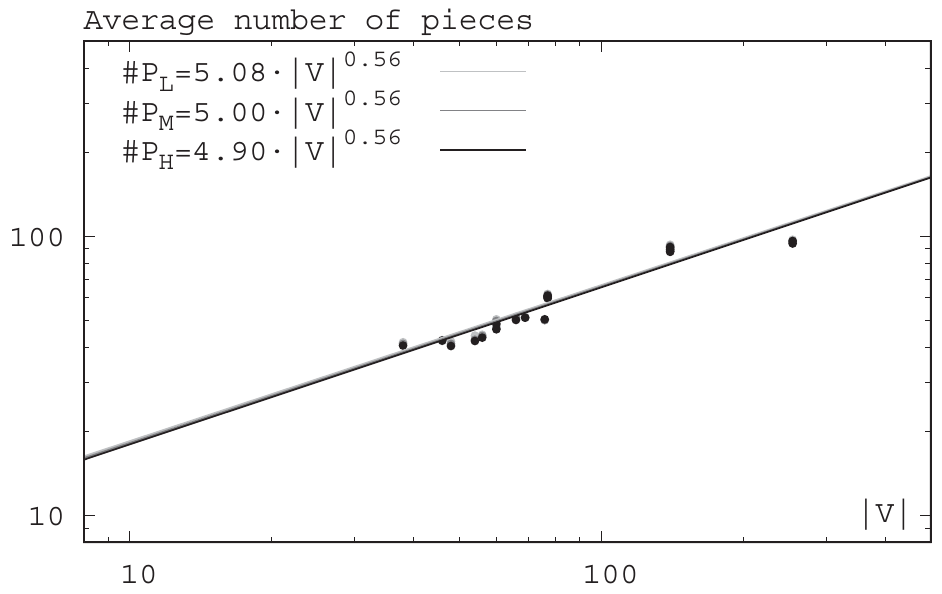}
    \end{minipage}
    \caption{Performance of the continuous quickest path algorithm for instances with low (L), medium (M) and high (H) time dependency}
    \label{fig:tdsp_running_time}
\end{figure}

We evaluate the performance of the continuous quickest-path algorithm presented in Section~\ref{sec_shortest_path}. Since this algorithm is only run during preprocessing, its results may be retained for successive routing solutions (e.g., for different customer sets) as long as the speed estimates are unchanged. Figure~\ref{fig:tdsp_running_time} reports the total CPU time spent by the algorithm as well as the average number of function pieces in the resulting $\Phi$ functions (over all origin-destination pairs) for each of the three instance categories. To evaluate the asymptotical behavior of the algorithm, we fitted these values as a power-law $f : |V| \rightarrow \alpha |V|^\beta$ via a least-squares regression of an affine function on the log-log graph.

Based on these results, we observe that the magnitude of the time dependency (L, M, and H) has only a limited impact on the performance of the quickest-path algorithm. The time needed to preprocess the continuous quickest-paths for all origin-destination pairs grows as $|V|^{x}$ with \mbox{$2.65 \leq x \leq 2.72$} and remains smaller than 40 seconds on all test instances. Moreover, the average number of function pieces for all origin-destination pairs only grows in $|V|^{0.56}$, reaching approximately 100 pieces in the largest instances. These average-case observations contrast with worst-case analyzes predicting an exponential growth \citep{Foschini2014}. To further reduce the preprocessing time whenever needed, the quickest-path algorithm could also be applied in parallel (calculating quickest-paths from different origins on different processors), or replaced by more sophisticated approaches based on contraction hierarchies \citep{Batz2013,Bast2016}.

\subsection{Performance of the Branch-Cut-and-Price Algorithm}
\label{sec_bcp}

Next, we evaluate the performance of the BCP algorithm. Tables~\ref{tbl:bcp-low}~to~\ref{tbl:bcp-high} present its computational performance for the low, medium, and high time dependency instances, excluding the largest two data sets from \cite{Brandao2008} with 347 and 375 services, for which the root-node CG could not be completed. The results are divided into two groups: Root Node and Branch-Cut-and-Price. In each group, columns LB, Gap, and T(s) respectively represent the lower bound, the percentage gap, and the time in seconds. For the complete BCP algorithm, the additional columns UB, N\textsubscript{E}, N\textsubscript{H}, R, and C represent the upper bound, the number of BCP nodes explored with exact pricing, the number of BCP nodes explored with the fast heuristic pricing during strong branching, the number of routes and cuts.
The percentage gaps are calculated as \mbox{100 $\times$ (UB -- LB) / LB}.
It is also important to observe that the BCP algorithm starts with the UB found by the metaheuristic, and that no further improvement on this solution was obtained for any instance. 

\begin{table}[htbp]
\renewcommand{\arraystretch}{1.15}
\setlength{\tabcolsep}{6pt}
\centering
\scalebox{0.8}{
\begin{tabular}{ccrrrrrrrrrrrr}\hline
& & \multicolumn{3}{c}{Root Node} &  & \multicolumn{8}{c}{Branch-Cut-and-Price} \\\cline{3-5}\cline{7-14}
Inst & $|E_R|$ & \multicolumn{1}{c}{LB} & \multicolumn{1}{c}{Gap} & \multicolumn{1}{c}{T(s)} &  & \multicolumn{1}{c}{LB} & \multicolumn{1}{c}{UB} & \multicolumn{1}{c}{Gap} & \multicolumn{1}{c}{T(s)} & \multicolumn{1}{c}{N\textsubscript{E}} & \multicolumn{1}{c}{N\textsubscript{H}} & \multicolumn{1}{c}{R} & \multicolumn{1}{c}{C} \\\hline
C01 &79& 2350.97 & 1.28\% & 55.26 && 2365.23 & 2381.01 & 0.67\% & 21599.70 & 163 & 4870 & 20074 & 748 \\
C02 &53& 1869.36 & 0.26\% & 13.16 && \textbf{1874.24} & \textbf{1874.24} & \textbf{0.00\%} & 131.24 & 15 & 432 & 4246 & 200 \\
C03 &51& 1592.70 & 0.60\% & 41.03 && \textbf{1602.26} & \textbf{1602.26} & \textbf{0.00\%} & 549.93 & 19 & 480 & 4956 & 211 \\
C04 &72& 1953.48 & 0.31\% & 131.38 && \textbf{1959.58} & \textbf{1959.58} & \textbf{0.00\%} & 4377.85 & 39 & 1191 & 12658 & 373 \\
C05 &65& 2485.97 & 0.94\% & 26.57 && \textbf{2509.43} & \textbf{2509.43} & \textbf{0.00\%} & 18320.00 & 183 & 5620 & 16306 & 596 \\
C06 &51& 1570.26 & 0.15\% & 29.11 && \textbf{1572.59} & \textbf{1572.59} & \textbf{0.00\%} & 371.37 & 13 & 393 & 6204 & 199 \\
C07 &52& 2058.63 & 0.58\% & 16.80 && \textbf{2070.49} & \textbf{2070.49} & \textbf{0.00\%} & 277.25 & 17 & 501 & 4165 & 345 \\
C08 &63& 1880.55 & 1.02\% & 21.27 && \textbf{1899.67} & \textbf{1899.67} & \textbf{0.00\%} & 2191.33 & 87 & 2596 & 8482 & 604 \\
C09 &97& 3329.41 & 0.63\% & 304.26 && 3345.64 & 3350.24 & 0.14\% & 21599.70 & 118 & 3586 & 17581 & 759 \\
C10 &55& 2126.94 & 1.71\% & 5.81 && \textbf{2163.26} & \textbf{2163.26} & \textbf{0.00\%} & 660.77 & 79 & 2317 & 5378 & 478 \\\hline
Avg C && 2121.83 & 0.75\% & 64.46 && 2136.24 & 2138.28 & 0.08\% & 7007.91 & 73 & 2199 & 10005 & 451 \\\hline
egl-e1 &51& 1618.94 & 0.56\% & 41.58 && \textbf{1627.97} & \textbf{1627.97} & \textbf{0.00\%} & 877.12 & 33 & 852 & 5049 & 244 \\
egl-e2 &72& 2327.27 & 0.67\% & 138.07 && 2339.52 & 2342.83 & 0.14\% & 21599.70 & 230 & 6711 & 11970 & 630 \\
egl-e3 &87& 2875.33 & 1.01\% & 255.83 && 2895.57 & 2904.36 & 0.30\% & 21599.70 & 191 & 5575 & 12574 & 811 \\
egl-e4 &98& 3305.87 & 1.07\% & 270.04 && 3323.42 & 3341.41 & 0.54\% & 21599.70 & 219 & 6383 & 11013 & 832 \\
egl-s1 &75& 2289.11 & 0.25\% & 256.43 && \textbf{2294.85} & \textbf{2294.85} & \textbf{0.00\%} & 2489.01 & 25 & 714 & 6352 & 414 \\
egl-s2 &147& 4361.29 & 0.49\% & 886.99 && 4369.29 & 4382.64 & 0.31\% & 21598.50 & 101 & 2991 & 10778 & 1986 \\
egl-s3 &159& 4608.44 & 0.55\% & 1750.46 && 4615.94 & 4633.79 & 0.39\% & 21598.50 & 59 & 1677 & 9489 & 903 \\
egl-s4 &190& 5470.11 & 1.21\% & 1573.58 && 5477.57 & 5536.26 & 1.07\% & 21598.40 & 62 & 1757 & 9900 & 1187 \\\hline
Avg egl && 3357.05 & 0.73\% & 646.62 && 3368.02 & 3383.01 & 0.34\% & 16620.08 & 115 & 3333 & 9641 & 876 \\\hline
\end{tabular}}
\caption{Performance of the BCP algorithm on low time-dependency instances.}
\label{tbl:bcp-low}
\end{table}

\begin{table}[htbp]
\renewcommand{\arraystretch}{1.15}
\setlength{\tabcolsep}{6pt}
\centering
\scalebox{0.8}{
\begin{tabular}{ccrrrrrrrrrrrr}\hline
 && \multicolumn{3}{c}{Root Node} &  & \multicolumn{8}{c}{Branch-Cut-and-Price} \\\cline{3-5}\cline{7-14}
Inst & $|E_R|$ & \multicolumn{1}{c}{LB} & \multicolumn{1}{c}{Gap} & \multicolumn{1}{c}{T(s)} &  & \multicolumn{1}{c}{LB} & \multicolumn{1}{c}{UB} & \multicolumn{1}{c}{Gap} & \multicolumn{1}{c}{T(s)} & \multicolumn{1}{c}{N\textsubscript{E}} & \multicolumn{1}{c}{N\textsubscript{H}} & \multicolumn{1}{c}{R} & \multicolumn{1}{c}{C} \\\hline
C01 &79& 2401.73 & 0.92\% & 92.07 && 2417.71 & 2423.80 & 0.25\% & 21599.70 & 169 & 5009 & 18296 & 807 \\
C02 &53& 1882.00 & 0.75\% & 13.90 && \textbf{1896.14} & \textbf{1896.14} & \textbf{0.00\%} & 1049.45 & 69 & 2129 & 7879 & 367 \\
C03 &51& 1626.26 & 0.71\% & 67.76 && \textbf{1637.80} & \textbf{1637.80} & \textbf{0.00\%} & 1082.72 & 27 & 762 & 5768 & 209 \\
C04 &72& 1986.64 & 0.51\% & 110.18 && \textbf{1996.75} & \textbf{1996.75} & \textbf{0.00\%} & 5089.53 & 53 & 1538 & 13986 & 427 \\
C05 &65& 2504.32 & 0.57\% & 21.58 && \textbf{2518.55} & \textbf{2518.55} & \textbf{0.00\%} & 199.43 & 17 & 478 & 3630 & 258 \\
C06 &51& 1581.11 & 0.06\% & 35.10 && \textbf{1582.08} & \textbf{1582.08} & \textbf{0.00\%} & 100.40 & 5 & 120 & 3483 & 149 \\
C07 &52& 2095.72 & 1.09\% & 18.62 && \textbf{2118.61} & \textbf{2118.61} & \textbf{0.00\%} & 822.74 & 55 & 1595 & 5980 & 427 \\
C08 &63& 1922.56 & 1.18\% & 17.48 && \textbf{1945.23} & \textbf{1945.23} & \textbf{0.00\%} & 1975.13 & 97 & 2822 & 7438 & 562 \\
C09 &97& 3368.35 & 1.16\% & 507.60 && 3389.81 & 3407.54 & 0.52\% & 21599.70 & 130 & 3932 & 15574 & 742 \\
C10 &55& 2163.62 & 2.03\% & 5.62 && \textbf{2207.63} & \textbf{2207.63} & \textbf{0.00\%} & 1472.44 & 149 & 4482 & 6730 & 555 \\\hline
Avg C && 2153.23 & 0.90\% & 88.99 && 2171.03 & 2173.41 & 0.08\% & 5499.12 & 77 & 2287 & 8876 & 450 \\\hline
egl-e1 &51& 1645.35 & 0.48\% & 45.46 && \textbf{1653.19} & \textbf{1653.19} & \textbf{0.00\%} & 1446.37 & 41 & 1082 & 4811 & 252 \\
egl-e2 &72& 2374.09 & 0.79\% & 184.52 && 2388.03 & 2392.91 & 0.20\% & 21599.70 & 213 & 6396 & 10608 & 701 \\
egl-e3 &87& 2925.54 & 1.04\% & 222.93 && 2946.82 & 2956.03 & 0.31\% & 21599.60 & 215 & 6232 & 12588 & 584 \\
egl-e4 &98& 3415.99 & 1.03\% & 367.68 && 3430.98 & 3451.16 & 0.59\% & 21599.70 & 203 & 5939 & 10310 & 566 \\
egl-s1 &75& 2368.82 & 0.55\% & 328.14 && 2378.22 & 2381.80 & 0.15\% & 21598.60 & 177 & 5269 & 13931 & 653 \\
egl-s2 &147& 4490.65 & 0.53\% & 975.90 && 4501.01 & 4514.46 & 0.30\% & 21598.50 & 129 & 3825 & 10822 & 1330 \\
egl-s3 &159& 4746.62 & 0.58\% & 2405.09 && 4753.27 & 4774.16 & 0.44\% & 21598.40 & 37 & 1089 & 8630 & 935 \\
egl-s4 &190& 5606.52 & 1.32\% & 3525.89 && 5612.43 & 5680.67 & 1.22\% & 21598.40 & 47 & 1373 & 9877 & 1470 \\\hline
Avg egl && 3446.70 & 0.79\% & 1006.95 && 3457.99 & 3475.55 & 0.40\% & 19079.91 & 133 & 3901 & 10197 & 811 \\\hline
\end{tabular}}
\caption{Performance of the BCP algorithm on medium time-dependency instances.}
\label{tbl:bcp-med}
\end{table}

\begin{table}[htbp]
\renewcommand{\arraystretch}{1.15}
\setlength{\tabcolsep}{6pt}
\centering
\scalebox{0.8}{
\begin{tabular}{ccrrrrrrrrrrrr}\hline
 && \multicolumn{3}{c}{Root Node} &  & \multicolumn{8}{c}{Branch-Cut-and-Price} \\\cline{2-4}\cline{6-13}
Inst & $|E_R|$ & \multicolumn{1}{c}{LB} & \multicolumn{1}{c}{Gap} & \multicolumn{1}{c}{T(s)} &  & \multicolumn{1}{c}{LB} & \multicolumn{1}{c}{UB} & \multicolumn{1}{c}{Gap} & \multicolumn{1}{c}{T(s)} & \multicolumn{1}{c}{N\textsubscript{E}} & \multicolumn{1}{c}{N\textsubscript{H}} & \multicolumn{1}{c}{R} & \multicolumn{1}{c}{C} \\\hline
C01 &79& 2416.45 & 1.15\% & 82.77 && 2431.77 & 2444.24 & 0.51\% & 21599.80 & 213 & 6211 & 18006 & 701 \\
C02 &53& 1877.57 & 0.65\% & 24.79 && \textbf{1889.78} & \textbf{1889.78} & \textbf{0.00\%} & 281.57 & 29 & 915 & 4584 & 223 \\
C03 &51& 1637.28 & 0.62\% & 77.36 && \textbf{1647.38} & \textbf{1647.38} & \textbf{0.00\%} & 1400.63 & 27 & 773 & 5474 & 151 \\
C04 &72& 1988.62 & 1.20\% & 102.67 && 2010.97 & 2012.42 & 0.07\% & 21599.80 & 155 & 4470 & 21694 & 609 \\
C05 &65& 2484.81 & 0.83\% & 16.81 && \textbf{2505.37} & \textbf{2505.37} & \textbf{0.00\%} & 8776.53 & 253 & 7784 & 12179 & 509 \\
C06 &51& 1563.85 & 0.15\% & 44.35 && \textbf{1566.24} & \textbf{1566.24} & \textbf{0.00\%} & 225.20 & 9 & 216 & 3939 & 154 \\
C07 &52& 2124.90 & 1.34\% & 14.53 && \textbf{2153.38} & \textbf{2153.38} & \textbf{0.00\%} & 2118.14 & 123 & 3584 & 7313 & 395 \\
C08 &63& 1938.63 & 1.09\% & 17.82 && \textbf{1959.71} & \textbf{1959.71} & \textbf{0.00\%} & 4523.27 & 147 & 4303 & 9396 & 563 \\
C09 &97& 3362.46 & 1.74\% & 645.53 && 3384.05 & 3420.94 & 1.09\% & 21599.70 & 132 & 3962 & 13146 & 564 \\
C10 &55& 2178.21 & 2.29\% & 6.54 && \textbf{2228.00} & \textbf{2228.00} & \textbf{0.00\%} & 12399.90 & 361 & 10873 & 10910 & 821 \\\hline
Avg C && 2157.28 & 1.10\% & 103.32 && 2177.67 & 2182.74 & 0.17\% & 9452.45 & 145 & 4309 & 10664 & 469 \\\hline
egl-e1 &51& 1659.50 & 0.71\% & 53.88 && \textbf{1671.31} & \textbf{1671.31} & \textbf{0.00\%} & 1519.80 & 51 & 1401 & 4353 & 247 \\
egl-e2 &72& 2406.54 & 1.00\% & 176.47 && 2426.65 & 2430.55 & 0.16\% & 21599.60 & 236 & 7010 & 9232 & 504 \\
egl-e3 &87& 2950.62 & 0.95\% & 285.06 && 2969.42 & 2978.79 & 0.32\% & 21599.70 & 229 & 6693 & 10822 & 587 \\
egl-e4 &98& 3524.30 & 0.89\% & 601.84 && 3537.15 & 3555.56 & 0.52\% & 21599.60 & 133 & 3992 & 8941 & 575 \\
egl-s1 &75& 2402.08 & 0.88\% & 515.27 && 2417.04 & 2423.32 & 0.26\% & 21598.60 & 195 & 5727 & 10668 & 644 \\
egl-s2 &147& 4602.66 & 0.82\% & 1499.58 && 4613.70 & 4640.30 & 0.58\% & 21598.50 & 89 & 2665 & 9846 & 1073 \\
egl-s3 &159& 4808.35 & 0.66\% & 1717.51 && 4816.24 & 4840.15 & 0.50\% & 21598.50 & 84 & 2509 & 8636 & 965 \\
egl-s4 &190& 5694.69 & 1.35\% & 5239.66 && 5700.98 & 5771.63 & 1.24\% & 21598.50 & 37 & 1147 & 9706 & 1100 \\\hline
Avg egl && 3506.09 & 0.91\% & 1261.16 && 3519.06 & 3538.95 & 0.45\% & 19089.10 & 132 & 3893 & 9026 & 712 \\\hline
\end{tabular}}
\caption{Performance of the BCP algorithm on high time-dependency instances.}
\label{tbl:bcp-high}
\end{table}

These results show that the amount of time-dependency has some impact on the BCP algorithm. With low time-dependency, eight optimal solutions are obtained for the instances C01--C10 and two for the EGL instances, leading to ten optimal solutions overall out of the 20 instances. In comparison, nine and eight optimal solutions have been found for the medium and high time-dependency cases, respectively. A study of the relative gaps confirms our observations: for the low time-dependency case, an average relative gap of 0.74\% is achieved at the root node, with a final average gap of 0.26\%. In contrast, the root node gap increases to $0.85\%$ and $1.02\%$ for the medium and high time-dependency cases, and the final average gap rises to $0.28\%$ and $0.37\%$. We also noted that the number of non-dominated labels explored during pricing at the root node is 30\% larger, on average, on the instances of type H than on the instances of type L. Finally, only a single call to the exact pricing algorithm was needed for $98.37\%$ of the nodes in type-L instances (i.e., only to prove that the CG is completed), whereas this ratio drops to $77.61\%$ for type-H instances.

\paragraph{Sensitivity analysis.} Our BCP relies on new methodological components that were carefully tailored for the TDCARP. We evaluate the impact of two key strategies: the exact pricing using heuristic dominance (HD) and the backward pricing + completion bounds (CB). Therefore, we compare BCP with two alternative configurations named BCP-noHD and BCP-noCB in which these components were deactivated. Tables~\ref{tbl:cpu:comp}~and~\ref{tbl:gap:comp} analyze the computational time of the three methods on the 27 instances which can be solved to optimality, as well as their final optimality gap on the remaining open instances.

\begin{table}[htbp]
\renewcommand{\arraystretch}{1.25}
\parbox{.48\linewidth}{
\centering
\begin{tabular}{cccc}
\hline
T(s)&BCP&BCP-noHD&BCP-noCB\\
\hline
Median&1400.63&1637.62&3588.61\\
Average&2767.74&2965.61&5964.68\\
\# Better&---&20&27\\
\# Equal&---&0&0\\
\# Worse&---&7&0\\
\hline
\end{tabular}
\caption{Comparison: Computational time}
\label{tbl:cpu:comp}
}
\hfill
\parbox{.48\linewidth}
{
\setlength{\tabcolsep}{6.35pt}
\begin{tabular}{cccc}
\hline
Gap(\%)&BCP&BCP-noHD&BCP-noCB\\
\hline
Median&0.39&0.40&0.55\\
Average&0.48&0.48&0.60\\
\# Better&---&13&27\\
\# Equal&---&5&0\\
\# Worse&---&9&0\\
\hline
\end{tabular}
\caption{Comparison: Final gap}
\label{tbl:gap:comp}
}
\end{table}

As visible in this sensitivity analysis, the use of the exact pricing with heuristic dominance contributes to a reduction of computational effort on 20 instances out of 27. Its effectiveness remains still limited by the necessity of exact dominance to complete the pricing process.
In contrast, the contribution of the backward pricing and completion bounds is visible for all instances, as it directly impacts the computational efficiency of the exact dominance phase. This approach considerably speeds up the solution process, permitting a more thorough search until the time limit and a substantial reduction of the final gap on open instances.

\subsection{Performance of the Hybrid Genetic Search}
\label{sec_hgs}

To evaluate the performance of the HGS, we run it ten times with different seeds on all instances, and compare its solutions with the lower bounds and optimal solutions (LB/Opt) obtained by the BCP. Tables~\ref{tbl:hgs-low}~to~\ref{tbl:hgs-high} report the results of this experiment. The leftmost columns report the instance names, number of services, and the known LB/Opt solution. Known optimal solutions are highlighted in boldface. The next columns report the CPU time spent calculating the quickest-path information. Finally, the remaining columns report the average and best solution quality of the HGS over the ten runs, the gap between the average and best solution values of the HGS, the gap between the average solution values of the HGS and the LB/Opt, and finally the CPU time of the~HGS.

\begin{table}[htbp]
\setlength{\tabcolsep}{6pt}
\renewcommand{\arraystretch}{1.15}
\centering
\scalebox{0.8}{
\begin{tabular}{ccccccccccccc}\hline
&&&BCP&&SP&&\multicolumn{5}{c}{HGS}\\\cline{8-12}
Inst&$|E_R|$&&LB/Opt&&T(s)&&Best&Avg&Gap&Gap$_\textsc{LB}$&T(s)\\
\hline
C01&79&&2365.23&&0.87&&2381.01&2381.16&0.01\%&0.67\%&556.31\\
C02&53&&\textbf{1874.24}&&0.34&&\textbf{1874.24}&\textbf{1874.24}&0.00\%&0.00\%&119.40\\
C03&51&&\textbf{1602.26}&&0.33&&\textbf{1602.26}&\textbf{1602.26}&0.00\%&0.00\%&183.26\\
C04&72&&\textbf{1959.58}&&0.63&&\textbf{1959.58}&\textbf{1959.58}&0.00\%&0.00\%&256.15\\
C05&65&&\textbf{2509.43}&&0.49&&\textbf{2509.43}&\textbf{2509.43}&0.00\%&0.00\%&166.92\\
C06&51&&\textbf{1572.59}&&0.23&&\textbf{1572.59}&\textbf{1572.59}&0.00\%&0.00\%&282.54\\
C07&52&&\textbf{2070.49}&&0.47&&\textbf{2070.49}&\textbf{2070.49}&0.00\%&0.00\%&105.58\\
C08&63&&\textbf{1899.67}&&0.79&&\textbf{1899.67}&\textbf{1899.67}&0.00\%&0.00\%&129.31\\
C09&97&&3345.64&&1.18&&3350.24&3350.24&0.00\%&0.14\%&479.55\\
C10&55&&\textbf{2163.26}&&0.58&&\textbf{2163.26}&\textbf{2163.26}&0.00\%&0.00\%&104.42\\
\hline
Avg C&&&&&0.59&&&&0.00\%&0.08\%&238.34\\
\hline
egl-e1&51&&\textbf{1627.97}&&1.10&&\textbf{1627.97}&\textbf{1627.97}&0.00\%&0.00\%&149.81\\
egl-e2&72&&2339.52&&1.15&&2342.83&2342.83&0.00\%&0.14\%&346.51\\
egl-e3&87&&2895.57&&1.12&&2904.36&2906.18&0.06\%&0.37\%&423.22\\
egl-e4&98&&3323.42&&1.11&&3341.41&3341.67&0.01\%&0.55\%&694.42\\
egl-s1&75&&\textbf{2294.85}&&6.86&&\textbf{2294.85}&\textbf{2294.85}&0.00\%&0.00\%&298.37\\
egl-s2&147&&4369.29&&7.06&&4382.64&4383.58&0.02\%&0.33\%&1195.87\\
egl-s3&159&&4615.94&&6.70&&4633.70&4641.87&0.18\%&0.56\%&3181.89\\
egl-s4&190&&5477.57&&7.05&&5524.07&5547.14&0.42\%&1.27\%&3520.23\\
egl-g1&347&&---&&25.17&&5659.43&5667.57&0.14\%&---&3600.12\\
egl-g2&375&&---&&28.49&&6543.11&6571.46&0.43\%&---&3600.13\\
\hline
Avg egl&&&&&8.58&&&&0.13\%&&1701.06\\
\hline
\end{tabular}}
\caption{Performance of the HGS on low time-dependency instances.}
\label{tbl:hgs-low}
\end{table}

\begin{table}[htbp]
\setlength{\tabcolsep}{6pt}
\renewcommand{\arraystretch}{1.15}
\centering
\scalebox{0.8}{
\begin{tabular}{ccccccccccccc}\hline
&&&BCP&&SP&&\multicolumn{5}{c}{HGS}\\\cline{8-12}
Inst&$|E_R|$&&LB/Opt&&T(s)&&Best&Avg&Gap&Gap$_\textsc{LB}$&T(s)\\
\hline
C01&79&&2417.71&&0.93&&2423.80&2423.80&0.00\%&0.25\%&522.30\\
C02&53&&\textbf{1896.14}&&0.34&&\textbf{1896.14}&\textbf{1896.14}&0.00\%&0.00\%&140.61\\
C03&51&&\textbf{1637.80}&&0.32&&\textbf{1637.80}&\textbf{1637.80}&0.00\%&0.00\%&247.90\\
C04&72&&\textbf{1996.75}&&0.69&&\textbf{1996.75}&\textbf{1996.75}&0.00\%&0.00\%&378.12\\
C05&65&&\textbf{2518.55}&&0.51&&\textbf{2518.55}&2522.22&0.15\%&0.15\%&210.63\\
C06&51&&\textbf{1582.08}&&0.24&&\textbf{1582.08}&\textbf{1582.08}&0.00\%&0.00\%&323.53\\
C07&52&&\textbf{2118.61}&&0.42&&\textbf{2118.61}&\textbf{2118.61}&0.00\%&0.00\%&129.54\\
C08&63&&\textbf{1945.23}&&0.85&&\textbf{1945.23}&\textbf{1945.23}&0.00\%&0.00\%&182.00\\
C09&97&&3389.81&&1.22&&3407.54&3408.20&0.02\%&0.54\%&574.99\\
C10&55&&\textbf{2207.63}&&0.59&&\textbf{2207.63}&\textbf{2207.63}&0.00\%&0.00\%&132.52\\
\hline
Avg C&&&&&0.61&&&&0.02\%&0.09\%&284.21\\
\hline
egl-e1&51&&\textbf{1653.19}&&1.17&&\textbf{1653.19}&\textbf{1653.19}&0.00\%&0.00\%&211.29\\
egl-e2&72&&2388.03&&1.47&&2392.91&2392.91&0.00\%&0.20\%&443.84\\
egl-e3&87&&2946.82&&1.16&&2956.03&2957.30&0.04\%&0.36\%&811.01\\
egl-e4&98&&3430.98&&1.18&&3451.16&3451.35&0.01\%&0.59\%&873.82\\
egl-s1&75&&2378.22&&7.07&&2381.80&2382.21&0.02\%&0.17\%&846.49\\
egl-s2&147&&4501.01&&6.95&&4514.46&4518.89&0.10\%&0.40\%&1696.38\\
egl-s3&159&&4753.27&&7.03&&4774.16&4784.27&0.21\%&0.65\%&2945.21\\
egl-s4&190&&5612.43&&7.22&&5663.44&5684.34&0.37\%&1.28\%&3431.32\\
egl-g1&347&&---&&26.54&&5776.54&5794.02&0.30\%&---&3600.16\\
egl-g2&375&&---&&27.50&&6702.28&6743.45&0.61\%&---&3600.17\\
\hline
Avg egl&&&&&8.73&&&&0.17\%&&1845.97\\
\hline
\end{tabular}}
\caption{Performance of the HGS on medium time-dependency instances.}
\label{tbl:hgs-med}
\end{table}

\begin{table}[htbp]
\setlength{\tabcolsep}{6pt}
\renewcommand{\arraystretch}{1.15}
\centering
\scalebox{0.8}{
\begin{tabular}{ccccccccccccc}\hline
&&&BCP&&SP&&\multicolumn{5}{c}{HGS}\\\cline{8-12}
Inst&$|E_R|$&&LB/Opt&&T(s)&&Best&Avg&Gap&Gap$_\textsc{LB}$&T(s)\\
\hline
C01&79&&2431.77&&1.01&&2444.24&2444.63&0.02\%&0.53\%&937.36\\
C02&53&&\textbf{1889.78}&&0.35&&\textbf{1889.78}&\textbf{1889.78}&0.00\%&0.00\%&189.71\\
C03&51&&\textbf{1647.38}&&0.35&&\textbf{1647.38}&\textbf{1647.38}&0.00\%&0.00\%&398.39\\
C04&72&&2010.97&&0.68&&2012.42&2012.42&0.00\%&0.07\%&400.12\\
C05&65&&\textbf{2505.37}&&0.53&&\textbf{2505.37}&2505.78&0.02\%&0.02\%&242.49\\
C06&51&&\textbf{1566.24}&&0.24&&\textbf{1566.24}&\textbf{1566.24}&0.00\%&0.00\%&331.34\\
C07&52&&\textbf{2153.38}&&0.46&&\textbf{2153.38}&\textbf{2153.38}&0.00\%&0.00\%&217.34\\
C08&63&&\textbf{1959.71}&&0.88&&\textbf{1959.71}&\textbf{1959.71}&0.00\%&0.00\%&216.92\\
C09&97&&3384.05&&1.31&&3420.94&3421.60&0.02\%&1.11\%&927.25\\
C10&55&&\textbf{2228.00}&&0.62&&\textbf{2228.00}&\textbf{2228.00}&0.00\%&0.00\%&174.13\\
\hline
Avg C&&&&&0.64&&&&0.01\%&0.17\%&403.50\\
\hline
egl-e1&51&&\textbf{1671.31}&&1.27&&\textbf{1671.31}&\textbf{1671.31}&0.00\%&0.00\%&289.68\\
egl-e2&72&&2426.65&&1.30&&2430.55&2430.55&0.00\%&0.16\%&557.60\\
egl-e3&87&&2969.42&&1.27&&2978.79&2979.35&0.02\%&0.33\%&907.83\\
egl-e4&98&&3537.15&&1.27&&3555.56&3555.65&0.00\%&0.52\%&1082.23\\
egl-s1&75&&2417.04&&7.23&&2423.32&2423.32&0.00\%&0.26\%&754.46\\
egl-s2&147&&4613.70&&7.28&&4640.30&4653.51&0.28\%&0.86\%&2171.17\\
egl-s3&159&&4816.24&&7.43&&4840.15&4855.75&0.32\%&0.82\%&3461.12\\
egl-s4&190&&5700.98&&7.75&&5766.76&5784.47&0.31\%&1.46\%&3600.04\\
egl-g1&347&&---&&35.29&&5834.33&5849.39&0.26\%&---&3600.21\\
egl-g2&375&&---&&30.88&&6827.08&6846.86&0.29\%&---&3600.26\\
\hline
Avg egl&&&&&10.10&&&&0.15\%&&2002.46\\
\hline
\end{tabular}}
\caption{Performance of the HGS on high time-dependency instances.}
\label{tbl:hgs-high}
\end{table}

As visible in these experiments, the HGS produces solutions of consistently good quality within a limited computational effort. 26 out of 27 optimal solutions have been attained on every run, whereas the optimal solution of the remaining instance (M-C05) has been attained on six runs out of ten. For the remaining instances which have no known optimum, the HGS finds average solutions that are guaranteed to be no further than 1.46\% from the optimum, based on the available lower bounds. The gap between average and best solutions over ten runs amounts to 0.08\% on average over all instances. This shows that the solution quality of the algorithm is stable over multiple executions. The HGS takes an average time of five minutes for the BMCV instances (C01--C10) and 31 minutes for the EGL instances. Faster runs could be achieved, if needed, by reducing the termination criterion, the population size, or by parallelizing some operations (e.g., generating multiple solutions in parallel). Finally, the level of time dependency has only a minor impact on computational performance: comparing the results on the instances with low (L) and high (H) time dependency, we observe a moderate increase of CPU time (by 24\%) but no significant deterioration of solution quality.

\begin{figure}[htbp]
    \hspace*{-0.5cm}
    \begin{minipage}{8.5cm}
    \includegraphics[width = \textwidth]{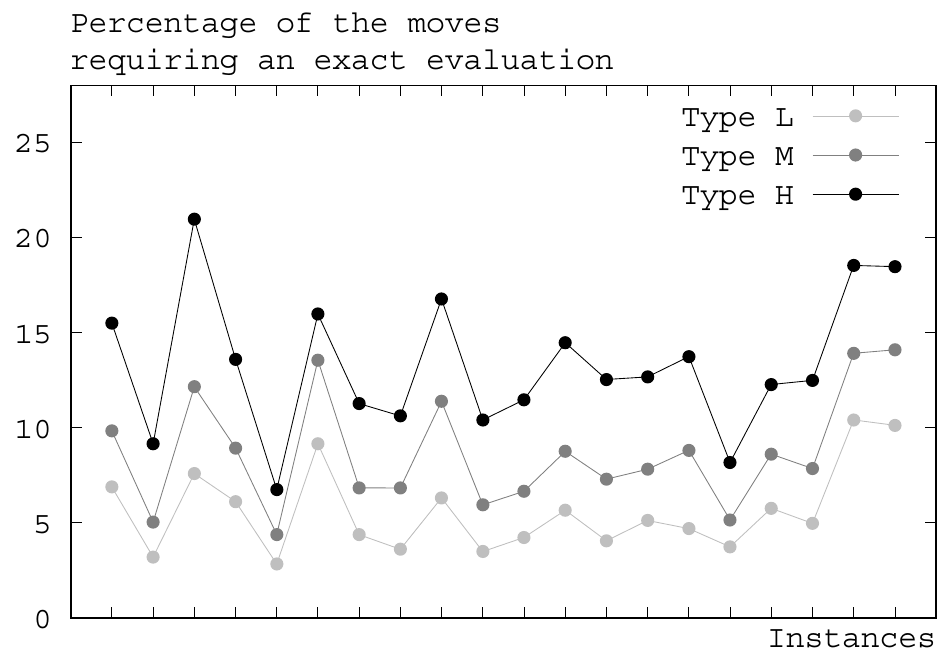}
    \end{minipage}
    \begin{minipage}{8.5cm}
    \includegraphics[width = \textwidth]{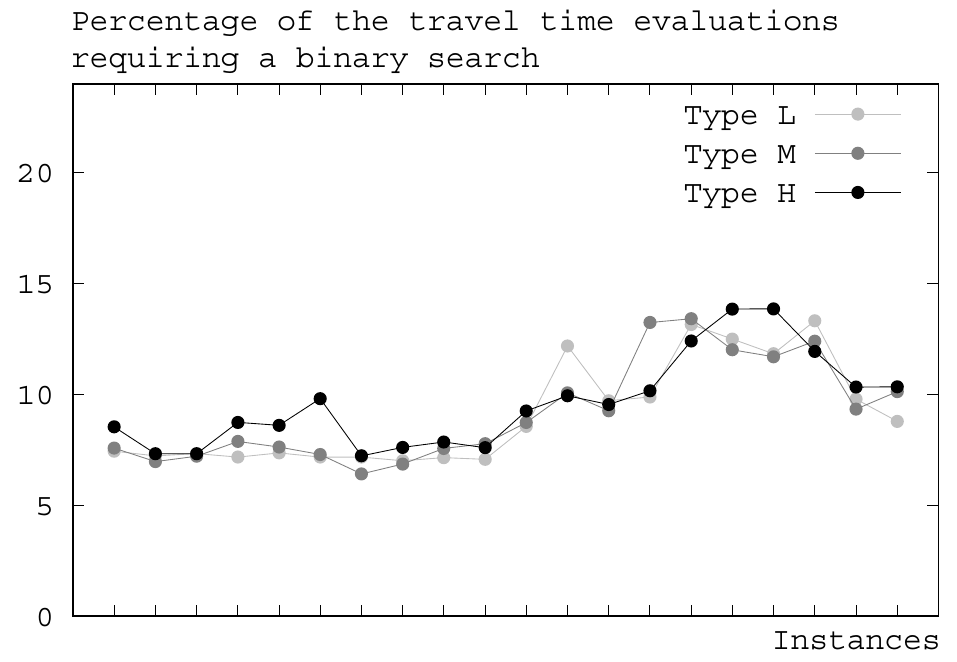}
    \end{minipage}
    \caption{Proportion of filtered moves and avoided binary searches in HGS}
    \label{fig:statsHGS}
\end{figure}

Next, we analyze the move lower bounds used to quickly filter non-improving moves (Equation~\ref{finalLBmove}) and the bucket structure used for fast travel times queries. Figure~\ref{fig:statsHGS} represents, for each time-dependency level and instance, the percentage of moves that needed exact evaluation, and the percentage of travel time queries that needed a binary search.

As visible in these results, Equation~\eqref{finalLBmove} allows to filter 91\% of the moves on average. The quality of the lower bounds depends on the amount of time dependency. For low time-dependency instances (type L), 96\% of the moves are filtered on average, whereas this number drops to 87\% for high time-dependency instances (type H). The bucket structures are also very effective, allowing to complete 90\% of the queries without binary search, regardless of the time dependency level.

\begin{figure}[htbp]
\centering
\includegraphics[width = 12.5cm]{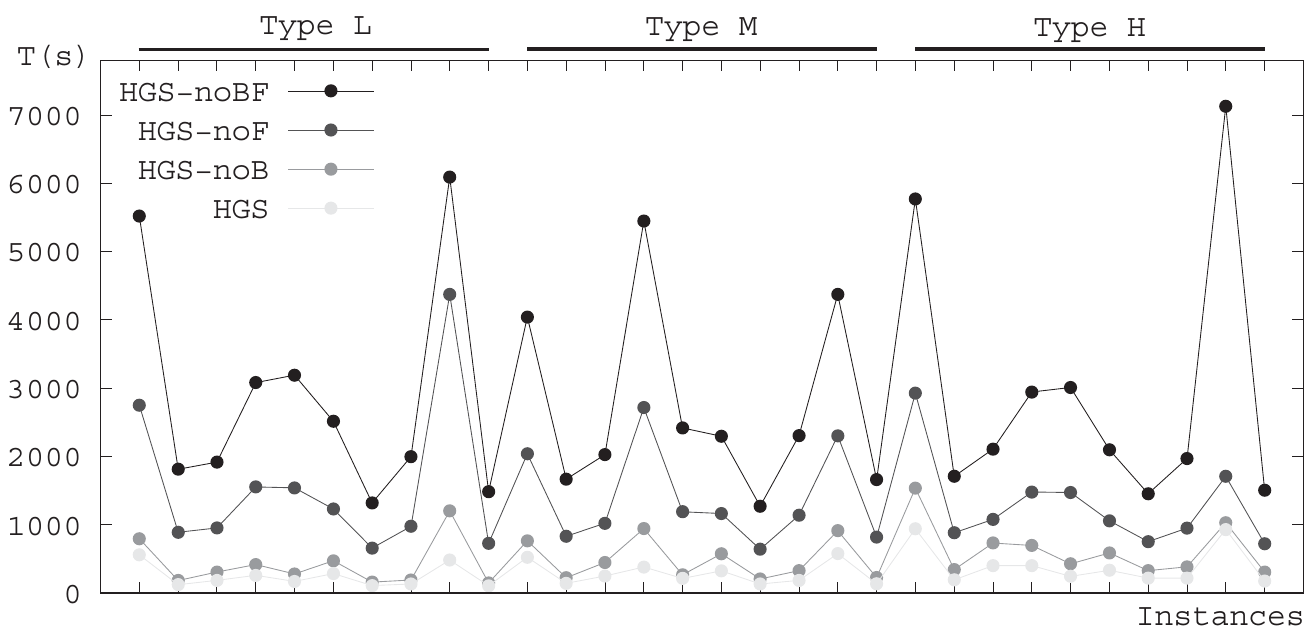}
\caption{Impact of the move filters and bucket-based function queries on the search time}
\label{fig:speedupsHGS}
\end{figure}

Both of the previously mentioned components contribute to speed up critical operations representing the bottleneck of the HGS. To highlight more directly their impact, Figure~\ref{fig:speedupsHGS} compares the CPU time of HGS on the instances C01--C10 with that of three alternative algorithm variants obtained by deactivating the move filters (\textsc{HGS-noF}), the bucket data structures (\textsc{HGS-noB}), or both of those (\textsc{HGS-noBF}). As shown by this experiment, the use of the move filters diminishes the computational effort by a factor five, while these two simple techniques jointly reduce the computational time by a factor ten on average, reaching up to $20$ for some instances. This time reduction allows to perform a much larger number of iterations within a reasonable time budget, therefore transforming speed-ups into quality gains.

\subsection{The Value of Time-dependent Route and Path Optimization}
\label{sec_value_of_tdcarp}

In a final experiment, we compare the solutions of the TDCARP and CARP algorithms in the presence of inaccurate speed information. For this analysis, we focus on the data sets C01--C10 for each time-dependency level (L, M, and H). For each of these 30 instances, we consider five levels of inaccuracy and generate each time $20$ travel-time scenarios by replacing the speed $v_{ij}(t)$ of each travel (and service) speed function piece by a value $v'_{ij}(t) = C v_{ij}(t)$, in which coefficient $C$ is sampled from a truncated normal distribution centered on $1$, with standard deviation $\sigma$ and truncation range $[1-\sigma,1+\sigma]$. Each level of inaccuracy corresponds to a different $\sigma$ in $\{0.05, 0.1, 0.2, 0.4, 0.6\}$.

\begin{figure}[htbp]
\centering
\begin{minipage}{4.5cm}
\centering
\includegraphics[width = 4.5cm]{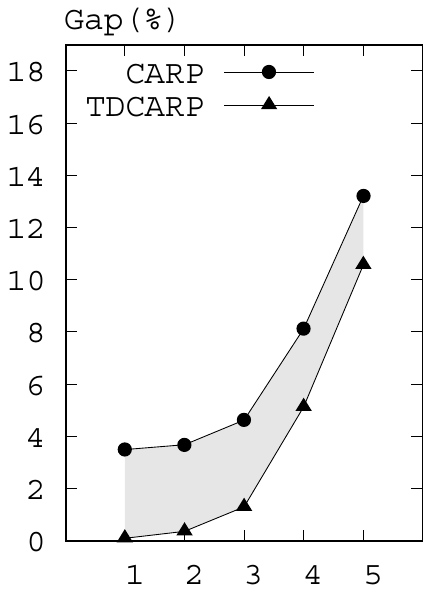}

\hspace*{0.5cm}Type L
\end{minipage}
\hspace*{0.8cm}
\begin{minipage}{4.5cm}
\centering
\includegraphics[width = 4.5cm]{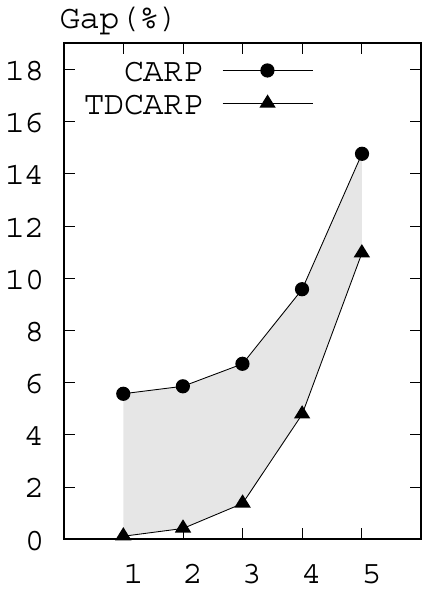}

\hspace*{0.5cm} Type M
\end{minipage}
\hspace*{0.8cm}
\begin{minipage}{4.5cm}
\centering
\includegraphics[width = 4.5cm]{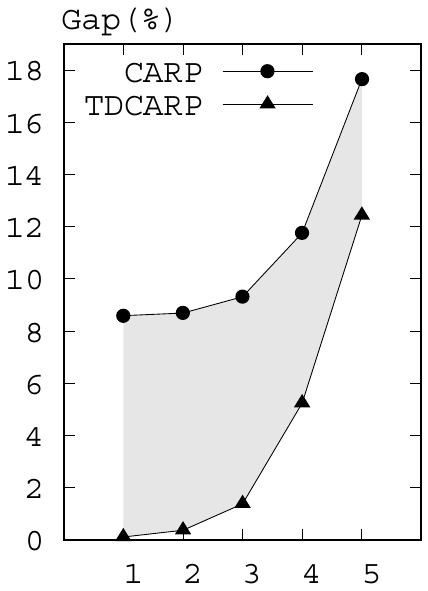}

\hspace*{0.5cm} Type H
\end{minipage}\vspace*{0.3cm}
\caption{The value of time-dependent routing, considering five level of speed information inaccuracy}
\label{fig:sentivity-information}
\end{figure}

We assume that only the original speed information $v_{ij}$ is available to the TDCARP optimization algorithm, whereas the CARP algorithm assumes uniform speed in the network. We measure the average quality of the solutions produced by the two methods over the $20$ randomized scenarios. This solution quality is then translated into an average error gap relative to the baseline TDCARP solutions obtained by running HGS with perfect knowledge of the speed conditions $v'_{ij}$ on each scenario. Figure~\ref{fig:sentivity-information} displays the result of this experiment, illustrating the average gap of the two methods on the five groups of scenarios and three instance types. The gap between the CARP and TDCARP solutions is grayed-out on the figure. It represents the value of time-dependent optimization.

As confirmed by these experiments, the accuracy of the speed information has an impact on TDCARP solutions. Inaccurate speed profiles naturally lead to larger gaps relative to the baseline. The grayed-out regions of~Figure~\ref{fig:sentivity-information} represent possible improvements that can be achieved by time-dependent routing optimization, whereas the regions below the TDCARP curve represent possible improvements which can arise by improving speed data collection processes. 

It is noteworthy that TDCARP solutions always improve upon CARP solutions, even in settings with very inaccurate knowledge of travel speeds (scenario 5 with $\sigma = 0.6$). Indeed, even in those cases, imperfect speed estimates give a better approximation of real scenarios than uniform speeds. Small inaccuracies (e.g., configurations with $\sigma \leq 0.2$) do not reduce the edge of TDCARP optimization, with gaps below $1.38\%$ in all cases relatively to the baseline. In contrast, merely omitting time dependency and using average speeds instead leads to a dramatic increase of travel time, which amounts to 6.28\% on average for these scenarios, and rises up to 9.32\% in high time-dependency cases (type H, scenario 3). Overall, despite data inaccuracies, time-dependent optimization remains essential for a good operational performance.

\section{Conclusions and Perspectives}
\label{sec:conclusion}

Time-dependent travel times defined at the network level pose significant challenges for solution algorithms. They dramatically increase the (methodological and computational) complexity of travel-time queries and significantly weaken label dominance in column generation-based approaches. Nevertheless, we demonstrated that these challenges can be overcome with dedicated solution strategies, including solution-decoding algorithms, move filters, heuristic dominance, enhanced pricing, and completion bounds. These methodological components allows us to solve small- to medium size-problems to optimality, and to produce good heuristic solutions for larger instances with several hundred services.

Our experiments demonstrate that time-dependent optimization is essential to slash down routing costs, even in scenarios in which speed information is inaccurate. This observation encourages us to use time-dependent routing algorithms in a broader range of applications, including cases with less accurate data sources \citep{Bertsimas2019b}.

The research perspectives are numerous. 
Firstly, our solution evaluations with inaccurate speed assumed that the routes should be fully defined at the start of the operations. As also mentioned in \cite{Gendreau2015}, a meaningful research line concerns the investigation of stochastic or dynamic routing problem, in which the travel-speed information collected during the day is used to rearrange customer-visit orders. Secondly, time-dependency dramatically weakens dominance relationships in column generation based methods, now based on simultaneous time and cost comparisons. A similar issue arises when considering generalized vehicle routing problems with time windows in which travel time and cost are not directly proportional \citep{BenTicha2019}. New methodological contributions are needed for these problems: better dominance rules, reduced-cost fixing techniques and completion-bounds may help. Finally, although we focused on arc-routing settings, we remind that an adequate management of time-dependent travel times at the network level remains a key issue for the entire vehicle routing community, and that most of the solution strategies listed in this paper are extensible to this broader context.

\ACKNOWLEDGMENT{
This research has been partially supported by CAPES, CNPq [grants numbers 308528/2018-2, 313521/2017-4, 425962/2016-4], and FAPERJ [grant number E-26/202.790/2019] in Brazil, as well as the Vietnam National Foundation for Science and Technology Development (NAFOSTED) [grant number 102.99-2016.21]. This support is gratefully acknowledged.
}

\begin{APPENDICES}

\section{Standard Iterative Algorithm for Arrival-Time Queries}

Algorithms \ref{alg:calPhi} and \ref{alg:calPhiInverse} allow to calculate $\Phi_{ij}(t_i)$ and $\Phi^{-1}_{ij}(t_j)$. We denote $v_{ij}(t^-) = \lim_{x \rightarrow t^-} v_{ij}(x)$ and $v_{ij}(t^+) = \lim_{x \rightarrow t^+} v_{ij}(x)$. These algorithms have been commonly used in the time-dependent vehicle routing literature \citep{Ichoua2003}. Their worst case complexity grows linearly with the number of pieces in the speed functions (in $\cO(h_{ij})$).

\begin{figure}[htbp]
\begin{minipage}{0.48\linewidth}
\begin{algorithm}[H]
\SingleSpacedXI
\small
\caption{Calculation of $\Phi_{ij}(t_i)$}
\label{alg:calPhi}
$t \gets t_i$ \;
$d \gets d_{ij}$\;
$k \gets \argmin \{ x \ | \ t_x > t_i \}$ \;
$t_j \gets t + d/v_{ij}(t^+)$ \;
\While {$t_j > t_k$} {
$d \gets d - v_{ij}(t^+) \times (t_k-t)$ \;
$t \gets t_k$ \;
$k \gets k+1$ \;
$t_j \gets t + d/v_{ij}(t^+)$ \;
}
\Return $t_j$ \;
\end{algorithm}
\end{minipage}
\hspace*{0.8cm}
\begin{minipage}{0.48\linewidth}
\begin{algorithm}[H]
\SingleSpacedXI
\small
\caption{Calculation of $\Phi^{-1}_{ij}(t_j)$}
\label{alg:calPhiInverse}
$t \gets t_j$  \;
$d \gets d_{ij}$  \;
$k \gets \argmax \{ x \ | \ t_x < t_j \}$ \;
$t_i \gets t - d/v_{ij}(t^-)$ \;
\While {$t_i < t_k$} {
    $d \gets d - v_{ij}(t^-) \times (t-t_k)$ \;
    $t \gets t_k$ \;
    $k \gets k-1$  \;
    $t_i \gets t - d/v_{ij}(t^-)$ \;
 }
 \Return $t_i$ \;
\end{algorithm}
\end{minipage}
\end{figure}

\section{Title Appendix B}
\label{sec:appB}

In the following examples, we demonstrate that time differences can arbitrarily diminish over a path. Consider two partial paths $P_1$ and $P_2$ currently finishing on the same service arc $(1,2)$ with $\Phi(P_1) = 1$ and $\Phi(P_2) = 5$. Both paths are extended on a sequence of three additional services $(2,3)  \rightarrow (3,4) \rightarrow (4,5)$ with service-and-travel speed functions defined as:
\begin{equation*}
\hat{v}_{23}(t) = \begin{cases} 1 &\text{ if } t \in [0,5] \\ 2 &\text{ otherwise }\end{cases} \quad \quad
\hat{v}_{34}(t) = \begin{cases} 1 &\text{ if } t \in [0,7] \\ 2 &\text{ otherwise }\end{cases} \quad \quad
\hat{v}_{45}(t) = \begin{cases} 1 &\text{ if } t \in [0,8] \\ 2 &\text{ otherwise }\end{cases}
\end{equation*}
With these parameters, label $P_1$ will complete the services to $(2,3)$, $(3,4)$  and $(4,5)$ at time $5$, $7$ and $8$ respectively, whereas label $P_2$ will complete the same services at time $7$, $8$ and $8.5$. The original time difference of $\Delta = 4$ units between the two labels has been reduced to $0.5$ units only after three services, and a trivial generalization of this example shows that it can diminish to $\Delta \left( \frac{v_\textsc{min}}{v_\textsc{max}} \right)^x$ after $x$ service or travel links, where $v_\textsc{min}$ and $v_\textsc{max}$ are the minimum and maximum possible speed values (here $v_\textsc{min}=1$ and $v_\textsc{max}=2$). 

\end{APPENDICES}


\begin{thebibliography}{44}
\expandafter\ifx\csname natexlab\endcsname\relax\def\natexlab#1{#1}\fi
\expandafter\ifx\csname url\endcsname\relax
  \def\url#1{{\tt #1}}\fi
\expandafter\ifx\csname urlprefix\endcsname\relax\def\urlprefix{URL }\fi
\expandafter\ifx\csname urlstyle\endcsname\relax
  \expandafter\ifx\csname doi\endcsname\relax
  \def\doi#1{doi:\discretionary{}{}{}#1}\fi \else
  \expandafter\ifx\csname doi\endcsname\relax
  \def\doi{doi:\discretionary{}{}{}\begingroup \urlstyle{rm}\Url}\fi \fi

\bibitem[{Baldacci et~al.(2011)Baldacci, Mingozzi, and Roberti}]{Roberti2011}
Baldacci, R., A.~Mingozzi, R.~Roberti. 2011.
\newblock New route relaxation and pricing strategies for the vehicle routing
  problem.
\newblock {\it Operations Research\/} {\bf 59}(5) 1269--1283.

\bibitem[{Bartolini et~al.(2013)Bartolini, Cordeau, and
  Laporte}]{Bartolini2013a}
Bartolini, E., J.-F. Cordeau, G.~Laporte. 2013.
\newblock Improved lower bounds and exact algorithm for the capacitated arc
  routing problem.
\newblock {\it Mathematical Programming\/} {\bf 137}(1) 409--452.

\bibitem[{Bast et~al.(2016)Bast, Delling, Goldberg, M{\"{u}}ller-Hannemann,
  Pajor, Sanders, Wagner, and Werneck}]{Bast2016}
Bast, H., D.~Delling, A.~Goldberg, M.~M{\"{u}}ller-Hannemann, T.~Pajor,
  P.~Sanders, D.~Wagner, R.F. Werneck. 2016.
\newblock {Route planning in transportation networks}.
\newblock L.~Kliemann, P.~Sanders, eds., {\it Algorithm Engineering: Selected
  Results and Surveys\/}. Springer, Berlin Heidelberg, 19--80.

\bibitem[{Batz et~al.(2013)Batz, Geisberger, Sanders, and Vetter}]{Batz2013}
Batz, G.V., R.~Geisberger, P.~Sanders, C.~Vetter. 2013.
\newblock {Minimum time-dependent travel times with contraction hierarchies}.
\newblock {\it Journal of Experimental Algorithmics\/} {\bf 18}(1) 1--43.

\bibitem[{Belenguer and Benavent(1998)}]{Belenguer1998}
Belenguer, J.M., E.~Benavent. 1998.
\newblock The capacitated arc routing problem: Valid inequalities and facets.
\newblock {\it Computational Optimization and Applications\/} {\bf 10}(2)
  165--187.

\bibitem[{{Ben Ticha} et~al.(2019){Ben Ticha}, Absi, Feillet, and
  Quilliot}]{BenTicha2019}
{Ben Ticha}, H., N.~Absi, D.~Feillet, A.~Quilliot. 2019.
\newblock {Multigraph modeling and adaptive large neighborhood search for the
  vehicle routing problem with time windows}.
\newblock {\it Computers {\&} Operations Research\/} {\bf 104}(1) 113--126.

\bibitem[{Bertsimas et~al.(2019)Bertsimas, Delarue, Jaillet, and
  Martin}]{Bertsimas2019b}
Bertsimas, D., A.~Delarue, P.~Jaillet, S.~Martin. 2019.
\newblock {Travel time estimation in the age of big data}.
\newblock {\it Operations Research\/} {\bf 67}(2) 498--515.

\bibitem[{Beullens et~al.(2003)Beullens, Muyldermans, Cattrysse, and {Van
  Oudheusden}}]{Beullens2003}
Beullens, P., L.~Muyldermans, D.~Cattrysse, D.~{Van Oudheusden}. 2003.
\newblock {A guided local search heuristic for the capacitated arc routing
  problem}.
\newblock {\it European Journal of Operational Research\/} {\bf 147}(3)
  629--643.

\bibitem[{Black et~al.(2013)Black, Eglese, and W{\o}hlk}]{Black2013}
Black, D., R.~Eglese, S.~W{\o}hlk. 2013.
\newblock {The time-dependent prize-collecting arc routing problem}.
\newblock {\it Computers {\&} Operations Research\/} {\bf 40}(2) 526--535.

\bibitem[{Brand{\~{a}}o and Eglese(2008)}]{Brandao2008}
Brand{\~{a}}o, J., R.~Eglese. 2008.
\newblock {A deterministic tabu search algorithm for the capacitated arc
  routing problem}.
\newblock {\it Computers {\&} Operations Research\/} {\bf 35}(4) 1112--1126.

\bibitem[{Cattaruzza et~al.(2017)Cattaruzza, Absi, Feillet, and
  Gonz{\'{a}}lez-Feliu}]{Cattaruzza2017}
Cattaruzza, D., N.~Absi, D.~Feillet, J.~Gonz{\'{a}}lez-Feliu. 2017.
\newblock {Vehicle routing problems for city logistics}.
\newblock {\it EURO Journal on Transportation and Logistics\/} {\bf 6}(1)
  51--79.

\bibitem[{Cookson and Pishue(2018)}]{Cookson2018}
Cookson, G., B.~Pishue. 2018.
\newblock {INRIX Global Traffic Scorecard}.
\newblock Tech. rep., INRIX Research.

\bibitem[{Corber{\'{a}}n and Laporte(2015)}]{Corberan2015}
Corber{\'{a}}n, {\'{A}}, G~Laporte. 2015.
\newblock {\it {Arc routing: Problems, methods, and applications}\/}.
\newblock Society for Industrial and Applied Mathematics, Philadelphia, PA,
  USA.

\bibitem[{Eglese et~al.(2006)Eglese, Maden, and Slater}]{Eglese2006}
Eglese, R., W.~Maden, A.~Slater. 2006.
\newblock {A road timetable to aid vehicle routing and scheduling}.
\newblock {\it Computers {\&} Operations Research\/} {\bf 33}(12) 3508--3519.

\bibitem[{Foschini et~al.(2014)Foschini, Hershberger, and Suri}]{Foschini2014}
Foschini, L., J.~Hershberger, S.~Suri. 2014.
\newblock {On the complexity of time-dependent shortest paths}.
\newblock {\it Algorithmica\/} {\bf 68}(4) 1075--1097.

\bibitem[{Gendreau et~al.(2015)Gendreau, Ghiani, and Guerriero}]{Gendreau2015}
Gendreau, M., G.~Ghiani, E.~Guerriero. 2015.
\newblock {Time-dependent routing problems: A review}.
\newblock {\it Computers {\&} Operations Research\/} {\bf 64} 189--197.

\bibitem[{Ghiani and Guerriero(2014)}]{Ghiani2014}
Ghiani, G., E.~Guerriero. 2014.
\newblock {A note on the Ichoua, Gendreau, and Potvin (2003) travel time
  model}.
\newblock {\it Transportation Science\/} {\bf 48}(3) 458--462.

\bibitem[{Hershberger(1989)}]{Hershberger1989}
Hershberger, J. 1989.
\newblock {Finding the upper envelope of n line segments in O(n log n) time}.
\newblock {\it Information Processing Letters\/} {\bf 33}(4) 169--174.

\bibitem[{Herszterg et~al.(2019)Herszterg, Poggi, and Vidal}]{Herszterg2019}
Herszterg, I., M.~Poggi, T.~Vidal. 2019.
\newblock {Two-dimensional phase unwrapping via balanced spanning forests}.
\newblock {\it INFORMS Journal on Computing\/} {\bf 31}(3) 527--543.

\bibitem[{Huang et~al.(2017)Huang, Zhao, {Van Woensel}, and Gross}]{Huang2017}
Huang, Y., L.~Zhao, T.~{Van Woensel}, J.-P. Gross. 2017.
\newblock {Time-dependent vehicle routing problem with path flexibility}.
\newblock {\it Transportation Research Part B: Methodological\/} {\bf 95}
  169--195.

\bibitem[{Ichoua et~al.(2003)Ichoua, Gendreau, and Potvin}]{Ichoua2003}
Ichoua, S., M.~Gendreau, J.-Y. Potvin. 2003.
\newblock {Vehicle dispatching with time-dependent travel times}.
\newblock {\it European Journal of Operational Research\/} {\bf 144}(2)
  379--396.

\bibitem[{Irnich(2008)}]{Irnich2008}
Irnich, S. 2008.
\newblock {Solution of real-world postman problems}.
\newblock {\it European Journal of Operational Research\/} {\bf 190}(1) 52--67.

\bibitem[{Jaballah et~al.(2019)Jaballah, Veenstra, Coelho, and
  Renaud}]{Jaballah2019}
Jaballah, R., M.~Veenstra, L.C. Coelho, J.~Renaud. 2019.
\newblock {The time-dependent shortest path and vehicle routing problem}.
\newblock Tech. rep., CIRRELT-2019-12.

\bibitem[{Laporte et~al.(2014)Laporte, Ropke, and Vidal}]{Laporte2014a}
Laporte, G., S.~Ropke, T.~Vidal. 2014.
\newblock {Heuristics for the vehicle routing problem}.
\newblock P.~Toth, D.~Vigo, eds., {\it Vehicle Routing: Problems, Methods, and
  Applications\/}, chap.~4. Society for Industrial and Applied Mathematics,
  87--116.

\bibitem[{Lera-Romero et~al.(2020)Lera-Romero, Miranda~Bront, and
  Soulignac}]{Lera-Romero2020}
Lera-Romero, G., J.J. Miranda~Bront, F.J. Soulignac. 2020.
\newblock Linear edge costs and labeling algorithms: The case of the
  time-dependent vehicle routing problem with time windows.
\newblock {\it Networks, Forthcoming\/} .

\bibitem[{Li and Eglese(1996)}]{Li1996}
Li, L.Y.O., R.W. Eglese. 1996.
\newblock {An interactive algorithm for vehicle routeing for winter-gritting}.
\newblock {\it Journal of the Operational Research Society\/} {\bf 47}(2)
  217--228.

\bibitem[{Maden et~al.(2009)Maden, Eglese, and Black}]{Maden2009}
Maden, W., R.~Eglese, D.~Black. 2009.
\newblock {Vehicle routing and scheduling with time-varying data: A case
  study}.
\newblock {\it Journal of the Operational Research Society\/} {\bf 61}(3)
  515--522.

\bibitem[{Martinelli et~al.(2014)Martinelli, Pecin, and Poggi}]{Martinelli2014}
Martinelli, R., D.~Pecin, M.~Poggi. 2014.
\newblock Efficient elementary and restricted non-elementary route pricing.
\newblock {\it European Journal of Operational Research\/} {\bf 239}(1) 102 --
  111.

\bibitem[{Mecler et~al.(2019)Mecler, Subramanian, and Vidal}]{Mecler2019}
Mecler, J., A.~Subramanian, T.~Vidal. 2019.
\newblock {A simple and effective hybrid genetic search for the job sequencing
  and tool switching problem}.
\newblock Tech. rep., PUC-Rio, ArXiV:1910.10021.

\bibitem[{Orda and Rom(1990)}]{Orda1990}
Orda, A., R~Rom. 1990.
\newblock {Shortest-path and minimum-delay algorithms in networks with
  time-dependent edge-length}.
\newblock {\it Journal of the Association for Computing Machinery\/} {\bf
  37}(3) 607--625.

\bibitem[{Pecin and Uchoa(2019)}]{Pecin2019}
Pecin, D., E.~Uchoa. 2019.
\newblock Comparative analysis of capacitated arc routing formulations for
  designing a new branch-cut-and-price algorithm.
\newblock {\it Transportation Science\/} {\bf 53}(6) 1673--1694.

\bibitem[{Pessoa et~al.(2010)Pessoa, Uchoa, Poggi, and Rodrigues}]{Pessoa2010}
Pessoa, A., E.~Uchoa, M.~Poggi, R.~Rodrigues. 2010.
\newblock Exact algorithm over an arc-time-indexed formulation for parallel
  machine scheduling problems.
\newblock {\it Mathematical Programming Computation\/} {\bf 2}(3) 259--290.

\bibitem[{Rincon-Garcia et~al.(2018)Rincon-Garcia, Waterson, and
  Cherrett}]{Rincon2018}
Rincon-Garcia, N., B.J. Waterson, T.J. Cherrett. 2018.
\newblock {Requirements from vehicle routing software: Perspectives from
  literature, developers and the freight industry}.
\newblock {\it Transport Reviews\/} {\bf 38}(1) 117--138.

\bibitem[{R{\o}pke(2012)}]{Rokpe2012}
R{\o}pke, S. 2012.
\newblock Branching decisions in branch-and-cut-and-price algorithms for
  vehicle routing problems.
\newblock Presentation, International Workshop on Column Generation, June 13,
  Bromont, QC, Canada.

\bibitem[{Sun et~al.(2015)Sun, Meng, and Tan}]{Sun2015}
Sun, J., Y.~Meng, G.~Tan. 2015.
\newblock {An integer programming approach for the Chinese postman problem with
  time-dependent travel time}.
\newblock {\it Journal of Combinatorial Optimization\/} {\bf 29}(3) 565--588.

\bibitem[{Toffolo et~al.(2019)Toffolo, Vidal, and Wauters}]{Toffolo2019}
Toffolo, T.A.M., T.~Vidal, T.~Wauters. 2019.
\newblock {Heuristics for vehicle routing problems: Sequence or set
  optimization?}
\newblock {\it Computers {\&} Operations Research\/} {\bf 105} 118--131.

\bibitem[{Vidal(2017)}]{Vidal2017b}
Vidal, T. 2017.
\newblock {Node, edge, arc routing and turn penalties: Multiple problems -- One
  neighborhood extension}.
\newblock {\it Operations Research\/} {\bf 65}(4) 992--1010.

\bibitem[{Vidal et~al.(2015{\natexlab{a}})Vidal, Battarra, Subramanian, and
  Erdogan}]{Vidal2014b}
Vidal, T., M.~Battarra, A.~Subramanian, G.~Erdogan. 2015{\natexlab{a}}.
\newblock {Hybrid metaheuristics for the clustered vehicle routing problem}.
\newblock {\it Computers {\&} Operations Research\/} {\bf 58}(1) 87--99.

\bibitem[{Vidal et~al.(2013)Vidal, Crainic, Gendreau, and Prins}]{Vidal2012a}
Vidal, T., T.G. Crainic, M.~Gendreau, C.~Prins. 2013.
\newblock {Heuristics for multi-attribute vehicle routing problems: A survey
  and synthesis}.
\newblock {\it European Journal of Operational Research\/} {\bf 231}(1) 1--21.

\bibitem[{Vidal et~al.(2014)Vidal, Crainic, Gendreau, and Prins}]{Vidal2012b}
Vidal, T., T.G. Crainic, M.~Gendreau, C.~Prins. 2014.
\newblock {A unified solution framework for multi-attribute vehicle routing
  problems}.
\newblock {\it European Journal of Operational Research\/} {\bf 234}(3)
  658--673.

\bibitem[{Vidal et~al.(2015{\natexlab{b}})Vidal, Crainic, Gendreau, and
  Prins}]{Vidal2015b}
Vidal, T., T.G. Crainic, M.~Gendreau, C.~Prins. 2015{\natexlab{b}}.
\newblock {Timing problems and algorithms: Time decisions for sequences of
  activities}.
\newblock {\it Networks\/} {\bf 65}(2) 102--128.

\bibitem[{Vidal et~al.(2020)Vidal, Laporte, and Matl}]{Vidal2019}
Vidal, T., G.~Laporte, P.~Matl. 2020.
\newblock {A concise guide to existing and emerging vehicle routing problem
  variants}.
\newblock {\it European Journal of Operational Research, Forthcoming\/} .

\bibitem[{Vidal et~al.(2016)Vidal, Maculan, Ochi, and Penna}]{Vidal2014}
Vidal, T., N.~Maculan, L.S. Ochi, P.H.V. Penna. 2016.
\newblock {Large neighborhoods with implicit customer selection for vehicle
  routing problems with profits}.
\newblock {\it Transportation Science\/} {\bf 50}(2) 720--734.

\bibitem[{Yu and Lin(2015)}]{Yu2015}
Yu, V.F., S.-W. Lin. 2015.
\newblock {Iterated greedy heuristic for the time-dependent prize-collecting
  arc routing problem}.
\newblock {\it Computers and Industrial Engineering\/} {\bf 90} 54--66.

\end{thebibliography}

\end{document}